\documentclass[fleqn,twoside]{article}
\usepackage{espcrc2}
\usepackage{graphicx}
\usepackage{psfrag}
\usepackage{amsmath}
\newcommand{\figwidth}{2.8in}
\newcommand{\figwidthtwo}{2.5in}
\title{Design Constraints for a WIMP Dark Matter and $pp$ Solar Neutrino Liquid Neon Scintillation Detector}
\author{M.G. Boulay\thanks{corresponding author, email: mboulay@lanl.gov}, A. Hime, and J. Lidgard\\\vspace{0.1in}Physics Division\\MS H803, Los Alamos National Laboratory\\Los Alamos, NM, USA 87545}

\begin{document}

\begin{abstract}
Detailed Monte-Carlo simulations were used to evaluate the performance of a liquid neon scintillation detector
for dark matter and low-energy solar neutrino interactions. A maximum-likelihood event vertex fitter including PMT time information
was developed, which significantly improves position resolution over spatial-only
algorithms, and substantially decreases the required detector size and achievable analysis energy threshold. 
The ultimate sensitivity to WIMP dark matter and the $pp$ flux uncertainty are evaluated as a function of detector size. The dependence on the neon scintillation and PMT properties are evaluated.
  A 300 cm radius detector would allow a $\sim$13 keV threshold,  a $pp$ flux uncertainty of $\sim$1$\%$, and limits on the spin-independent WIMP-nucleon cross-section of
$\sim10^{-46}$ cm$^2$ for a 100 GeV WIMP, using commercially available PMTs.  Detector response calibration and background requirements for a precision $pp$ measurement are defined. Internal radioactivity requirements for uranium, thorium, and krypton
are specified, and it is shown that the PMT data could be used for an {\it{in-situ}} calibration of the troublesome ${}^{85}$Kr.
A set of measurements of neon scintillation properties and PMT characteristics are outlined which will be needed in order to evaluate feasibility and fully optimize the design of a neon-based detector.
\end{abstract}
\maketitle
\tableofcontents
\section{Introduction}
This paper presents details on the design of a dual-purpose liquid neon based detector for dark matter and low-energy solar neutrino interactions.
The origin of dark matter in our universe and study of low-energy solar neutrinos are two of the foremost topics in particle astrophysics.
Current observational evidence strongly suggests that a large fraction of the total matter in our universe is non-luminous, non-baryonic
matter outside of the current standard model of particle physics.  A currently favored hypothesis is that the dark matter consists of WIMPs (Weakly Interacting Massive Particles).
A natural candidate for this particle is the lightest superpartner which arises from supersymmetry,
the neutralino.  Current direct dark matter searches attempt to detect WIMP interactions through elastic scattering on target nuclei, and this is the basis for WIMP detection in this study.

The current generation of solar neutrino experiments have shown that mixing occurs between the neutrino flavors.  One of the goals for next-generation experiments is to measure with high-precision the low-energy $pp$ neutrinos.  These neutrinos account for greater than 90\% of all solar neutrinos, and are well-constrained by the measured solar luminosity.  A precision measurement of the flux of $pp$ neutrinos would lead to improved knowledge of neutrino mixing in the solar sector ($\theta_{12}$),  tests of the neutrino oscillation prediction in the low-energy region where vacuum mixing dominates, and precision tests of solar evolution models. 

WIMP particles ($\chi$) and solar neutrinos ($\nu$) can be observed in liquid neon through their elastic scattering from nuclei and electrons, respectively:
\begin{eqnarray}
\chi + {\rm{Ne}} &\longrightarrow& \chi^{'} + {\rm{Ne}}^{'} \label{wimp_scatter} \\
\nu + e^{-} & \longrightarrow & \nu^{'} + e^{-'} \label{nu_scatter}.
\end{eqnarray}
 Monte-Carlo simulations were performed to evaluate the performance of a neon scintillation detector sensitive to the recoiling nuclei (${\rm{Ne}}^{'}$) and electrons
($e^{-'}$) from equations~\ref{wimp_scatter} and \ref{nu_scatter}.
The concept of a liquid neon scintillation detector for solar neutrinos was first proposed by McKinsey and Doyle~\cite{clean_paper}.
The experimental design of a next-generation WIMP dark matter experiment should allow a sensitivity increase of several orders of magnitude over present dark matter experiments to probe interesting regions
of supersymmetry parameter space.  Ultra-low backgrounds and a large fiducial target are required to achieve this goal.  A large spherical volume of neon, along with precise event position reconstruction allows for a large fiducial mass and a large mass to surface area ratio, with the advantage that external source backgrounds are physically distant from the fiducial target and can be dealt with through position reconstruction.

Simulations were performed assuming nominal characteristics of scintillation light in liquid neon, and detector performance was estimated  for commercially available photomultiplier tubes (PMTs) with ultra-low background glass.  The details of the simulation inputs and algorithms are presented in section~2.
The detector response to scintillation events is characterized in section~3, where reconstruction algorithms and response functions useful for evaluating predicted observables for dark matter and solar
neutrino interactions are presented.
The sensitivity to solar neutrinos is discussed in section~4.  Section~5 describes the reduction of external and internal backgrounds through reconstruction 
and analysis cuts, and the ultimate background requirements.  Optimizing the detector for dark matter and solar neutrino sensitivity is discussed in section~6.
Calibration requirements for the full-scale experiment are outlined in section~7.  The requirements from photon absorption and sensitivity
to neon scintillation and detector parameters are presented in section~8, along with 
a list of which measurements are needed to fully optimize the detector design.  Conclusions from this work, and potential advantages of this approach are presented in section~9.

\section{Simulation Details}
A detailed simulation of a liquid neon detector was coded using the GEANT4 package~\cite{geant4}.  This package is used for tracking of electrons, $\gamma$ and x-ray photons, $\alpha$-particles, recoiling nuclei, and neutrons  in the detector, and subsequent simulation and tracking of scintillation photons.  A 3D model of the PMT is used, and tracking of photons across surfaces is accomplished
with the UNIFIED model of the DETECT code~\cite{unified_detect}, as incorporated in GEANT4.  All neutrino or dark matter interactions
are detected via scintillation light produced in the liquid neon.  Table~\ref{neon_scint_prop} shows the properties of liquid neon used in
the simulation, along with the references for these values.  
Scintillation light is produced via excitation of molecular states in the neon.  The relative intensity of these states, as well as the overall yield (or quenching factor) depends on the 
ionizing particle.  Two states are produced:  the singlet ($^{1}\Sigma$) state with an assumed 2.2 ns lifetime~\cite{singlet_lifetime} and the triplet ($^{3}\Sigma$) state with a 2.9 $\mu$s lifetime~\cite{triplet_lifetime}.

\begin{figure}[h]
\begin{center}
\includegraphics[width=\figwidth]{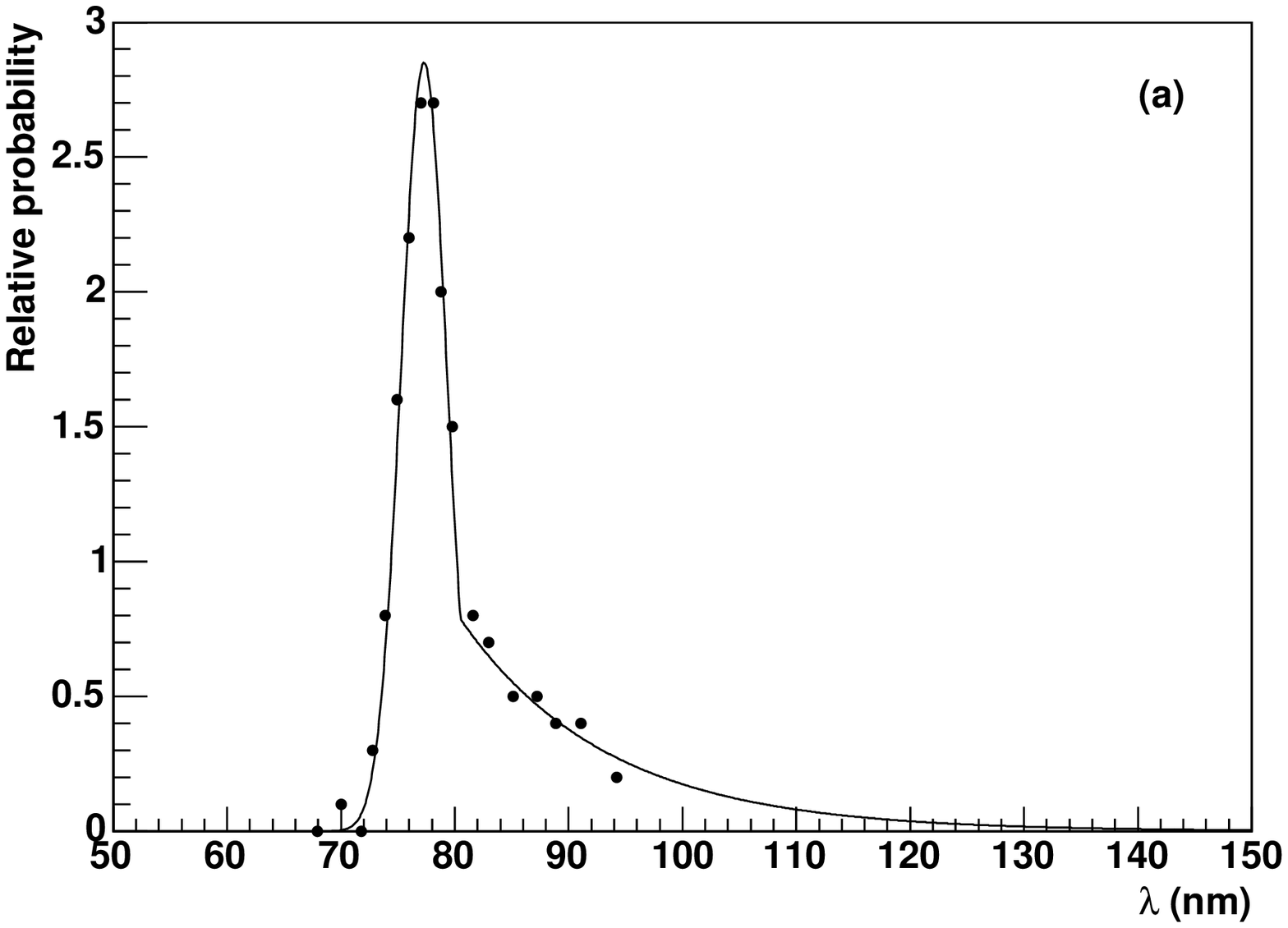}
\includegraphics[width=\figwidth]{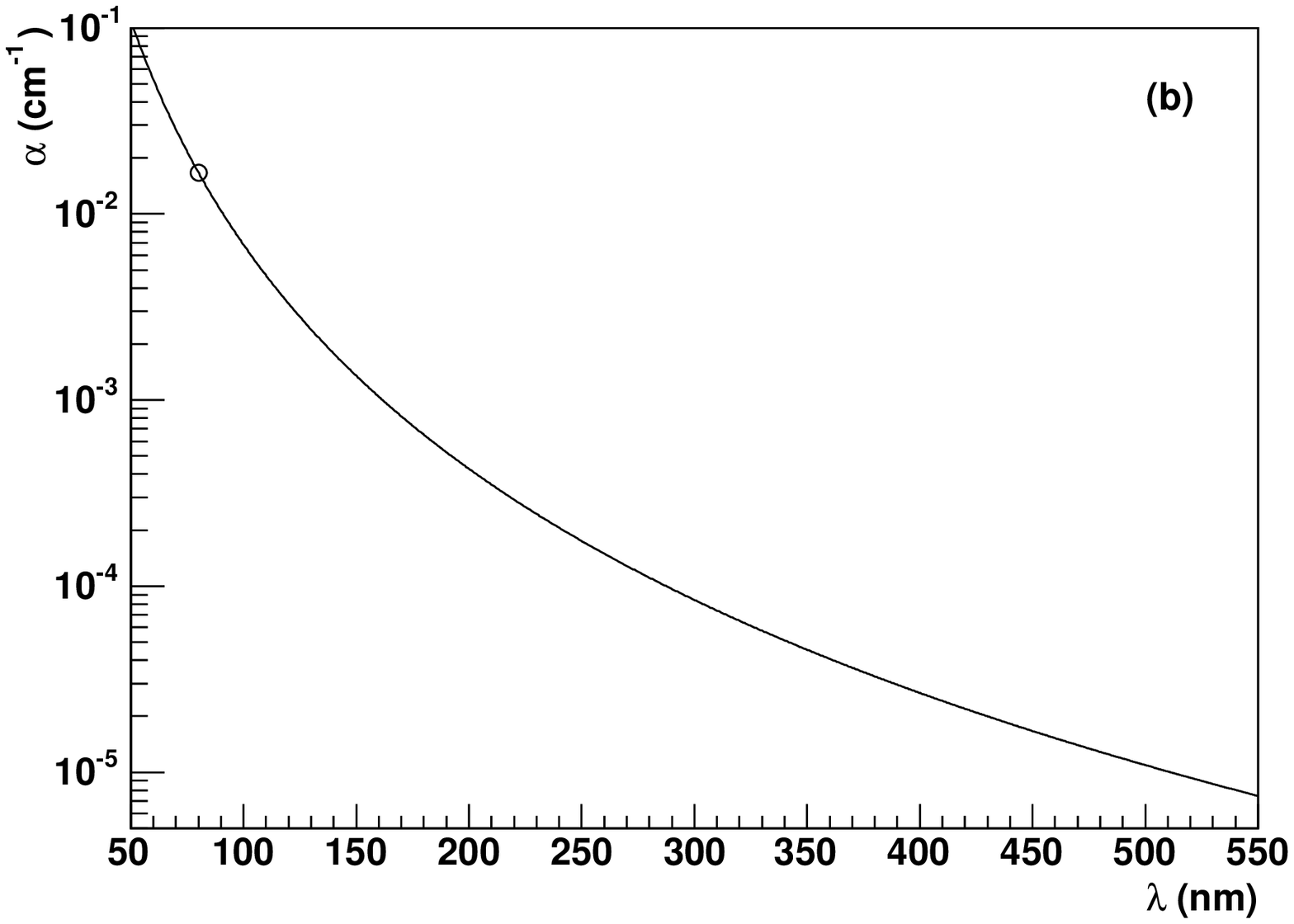}
\caption[Neon scintillation light]{\label{neon_scint_spectrum}(a) Wavelength spectrum of neon scintillation light. (b) Rayleigh scattering length in neon.  The Rayleigh scattering length versus wavelength has been extrapolated from the calculated value at 80 nm based on the $\lambda^{-4}$ dependence.}
\end{center}
\end{figure}

Quenching factors and the ratio of intensity of the singlet to triplet states ($^{1}I/^{3}I$), which determine the overall time structure of the scintillation light, for electrons, alpha particles, and nuclear recoils were inferred from a combination of measurements with liquid argon, xenon, and neon~\cite{hitachi,neon_scint} and are shown in Table~\ref{table_quench}.  The difference of these intensity ratios between electron and nuclear recoils provides discrimination between these event types and ultimately allows the separation of potential WIMP events from solar neutrino events and backgrounds.
Many of the entries in Tables~1 and~2 are estimates or preliminary measurements, and need to be quantified more precisely.
The sensitivity to these parameters is discussed in section~\ref{sensitivity_section}.
 To detect the scintillation photons with a 75\% photocathode coverage on a sphere with a nominal radius of R$_{\rm psup}$ = 300 cm (R$_{\rm psup}$ denotes the radius of the photomultipler support structure), 1832 12 cm hemispherical PMTs are arranged in a 3-frequency icosahedral pattern~\cite{icosahedron}. The nominal PMT properties  assumed in the simulation are shown in Table~\ref{pmt_properties}.  The commercially available Electron Tubes D737KB~\cite{electron_tubes} PMTs assumed for the simulation are made with ultra-low background glass, with  quoted upper limits of uranium and thorium of 30 parts per billion (ppb), and of potassium of 60 parts per million (ppm). These PMTs can be manufactured with low-resistivity photocathodes for operation at low temperatures~\cite{electron_tubes_lowt}.  The random count rate in the PMTs is assumed to be 500 Hz, extrapolated from 
measurements made by the manufacturer~\cite{electron_tubes_lowt_sheet}.
The spectrum of light emitted from scintillation in neon peaks at 77 nm, and is shown in Fig~\ref{neon_scint_spectrum}(a), adapted from~\cite{neon_spectrum}.  The Rayleigh scattering
length at this wavelength in liquid neon has been calculated to be~60 cm~\cite{lanou_rayleigh}.  Fig.~\ref{neon_scint_spectrum}(b) shows the Rayleigh scattering length versus photon
wavelength in liquid neon used in the simulation

\begin{figure}[h]
\begin{center}
\includegraphics[width=\figwidth]{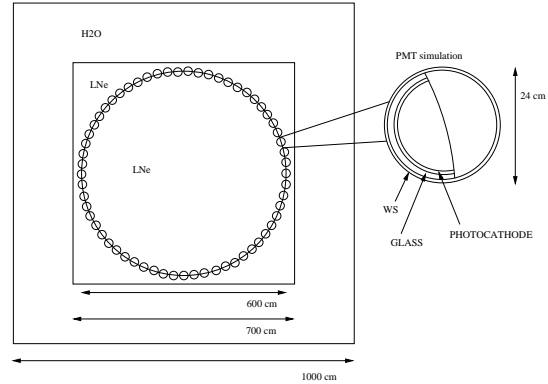}
\caption[Detector geometry]{\label{detector_geom}Detector geometry used in the simulation.  In the nominal configuration (R$_{\rm psup}$=300 cm), 1832 PMTs are placed at a radius of 3 meters in a 3-frequency icosahedral pattern, for 75\% coverage.}
\end{center}
\end{figure}

The nominal detector geometry used in the simulation is shown in Fig.~\ref{detector_geom}.  A sphere of PMTs is immersed in a volume of liquid neon.  
Each PMT views the inner region of neon, and consists of a hemispherical photocathode and glass window which has been covered with a wavelength shifting film.
Extreme ultraviolet (EUV) photons which strike the front of the PMT are shifted up in wavelength and are emitted into 4$\pi$.  These photons can then be transported through the window and generate a photoelectron 
in the cathode, according to the PMT quantum efficiency (0.20)  used in the calculation.  The wavelength shifter (WS) is assumed to be transparent to photons 
which have been previously shifted into the 250-500 nm range.

\begin{table*}[htb]
\begin{center}
\caption[Scintillation properties of liquid neon]{\label{neon_scint_prop}Properties of liquid neon used in the simulation.}
\begin{tabular}{lll} \hline
Quantity & Value & Reference \\ \hline
LNe scintillation yield (prompt)  & 10791.7 photons/MeV&\cite{neon_scint,helium_scint,lenou_taup,neon_scint_corr} \\
LNe scintillation spectrum & \small{(see Fig.~\ref{neon_scint_spectrum})}& \cite{neon_spectrum}\\ 
Rayleigh scattering length @ 80 nm& 60 cm& \cite{lanou_rayleigh}\\ 
refractive index in LNe @ 80 nm & 1.233 & \cite{lanou_rayleigh}\\ 
absorption length in LNe @ 80 nm & $\infty$& \\ 
singlet (prompt) time constant ($\tau_{S}$) & 2.2 ns&\cite{singlet_lifetime} \\ 
triplet (slow) time constant ($\tau_{T}$)&2.9 $\mu$s &\cite{triplet_lifetime} \\ 
prompt WS efficiency for EUV photons (4$\pi$) & 1.0 & \cite{waveshift}\\ 
total WS efficiency for EUV photons (4$\pi$) & 1.35& \cite{waveshift}\\ 
WS time constant \footnotemark[1] & 14.816 ns& \cite{waveshift}\\                
\multicolumn{3}{l}{\footnotemark[1] Time constant for 100$\%$ efficiency within 20 ns, inferred from~\cite{neon_scint}.} \\
\end{tabular}
\end{center}
\end{table*}

\vspace{0.4in}

\begin{table}[htb]
\begin{center}
\caption[Quenching factors used in simulation]{\label{table_quench}Quenching factors and singlet/triplet intensity ratios ($^{1}I/^{3}I$) for electrons, $\alpha$'s, and nuclear recoils (n.r.) used in the simulation.}
\begin{tabular}{llll} \hline
Property & e$^{-}$ & $\alpha$ & n.r. \\ \hline
quenching (prompt) & 1.0 &0.7973 & 0.25 \\
$^{1}I/^{3}I$  & 2.0 & 8.7 & 20 \\ \hline
\end{tabular}
\end{center}
\end{table}

\begin{table}[htb]
\begin{center}
\caption[PMT properties]{\label{pmt_properties}Nominal PMT properties assumed for simulation for R$_{\rm psup}$ = 300 cm.}
\begin{tabular}{ll} \hline
Property & Value \\ \hline
number of PMTs & 1832 \\ 
photo cathode radius & 12 cm \\ 
effective coverage & 0.75 \\ 
Q.E. (blue photons on envelope) & 0.20 \\ 
threshold & 1/4 p.e. \\
photo cathode reflectance & 0.02 \\ 
dark noise rate & 500 Hz \\ \hline
Electron Tubes D737KB & \\ \hline
mass (glass) & 1 kg \\
thickness (glass) & 3 mm \\
uranium & 30 ppb \\
thorium & 30 ppb \\
potassium & 60 ppm \\ \hline
\end{tabular}
\end{center}
\end{table}

\section{Detector response}
In this section the detector response functions in position and energy are evaluated using the Monte-Carlo calculations.  These
response functions can then be used to estimate the signal from WIMP and neutrino interactions, the backgrounds present in the signal
region, and the ultimate experimental sensitivity.
\subsection[Event reconstruction]{Event position and energy reconstruction}
Coakley and McKinsey have shown the potential for using completely geometrical position reconstruction algorithms for scintillation events
in a liquid neon detector~\cite{spatial_recon}.  Here we describe a more general reconstruction algorithm which incorporates PMT hit time as well as hit pattern information
to improve position resolution.  Although the PMT timing distribution is quite broad (many tens of ns), the distinctive shape of this
distribution for events near the center versus that near the PMT region improves the position resolution significantly.

In the absence of scattering or noise hits, the location of a scintillation event can be estimated by taking the geometric mean
of the PMT hit positions, since scintillation photons are emitted isotropically into $4\pi$:
\begin{eqnarray*}
\vec{R}_{event} &=& \frac{1}{N_{PMT}}\sum_{i=0}^{N_{PMT}} \vec{R}^{i}_{PMT} \\
                &\equiv& \vec{G}
\end{eqnarray*}
\noindent where $\vec{R}_{event}$ is the event location and $\vec{R}^{i}_{PMT}$ is the vector to the $i^{th}$ PMT to detect a scintillation photon.
Assuming spherical detector symmetry, the only effect of scattering can be a radial offset between the true event position and
the mean of the PMT hit positions, so that we have
\begin{equation}
\vec{R}_{event} = \vec{G} +\alpha \hat{r}  \label{r_est}
\end{equation}
\noindent where
\begin{equation}
\hat{r}  =  \frac{\vec{R}_{event}}{| \vec{R}_{event} | } \approx \frac{\vec{G}}{| \vec{G} | }
\end{equation}
\noindent and $\alpha$ is related to the effective scattering length in the detector.  Rearranging equation~\ref{r_est}, we have
\begin{equation}
\vec{R}_{event} = \vec{G}(1+\alpha)  \label{eqn:recon_cal},
\end{equation}
\noindent  absorbing an arbitrary constant and with $\alpha$ in general a function of event radius and energy.  We can thus determine $\alpha$ 
by Monte-Carlo and use equation~\ref{eqn:recon_cal} to estimate the position of scintillation events.

\begin{figure}[h]
\begin{center}
\includegraphics[width=\figwidth]{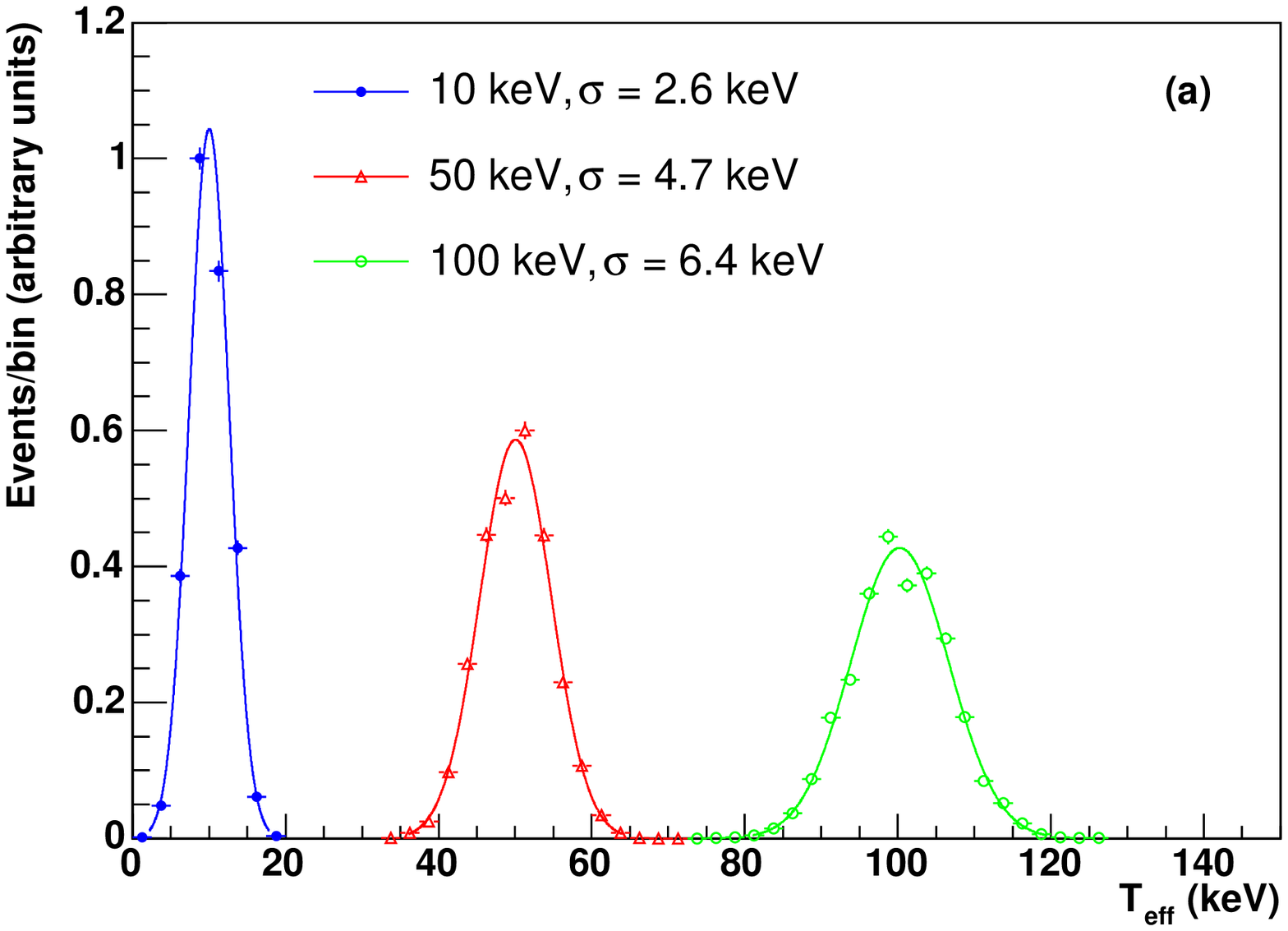}
\includegraphics[width=\figwidth]{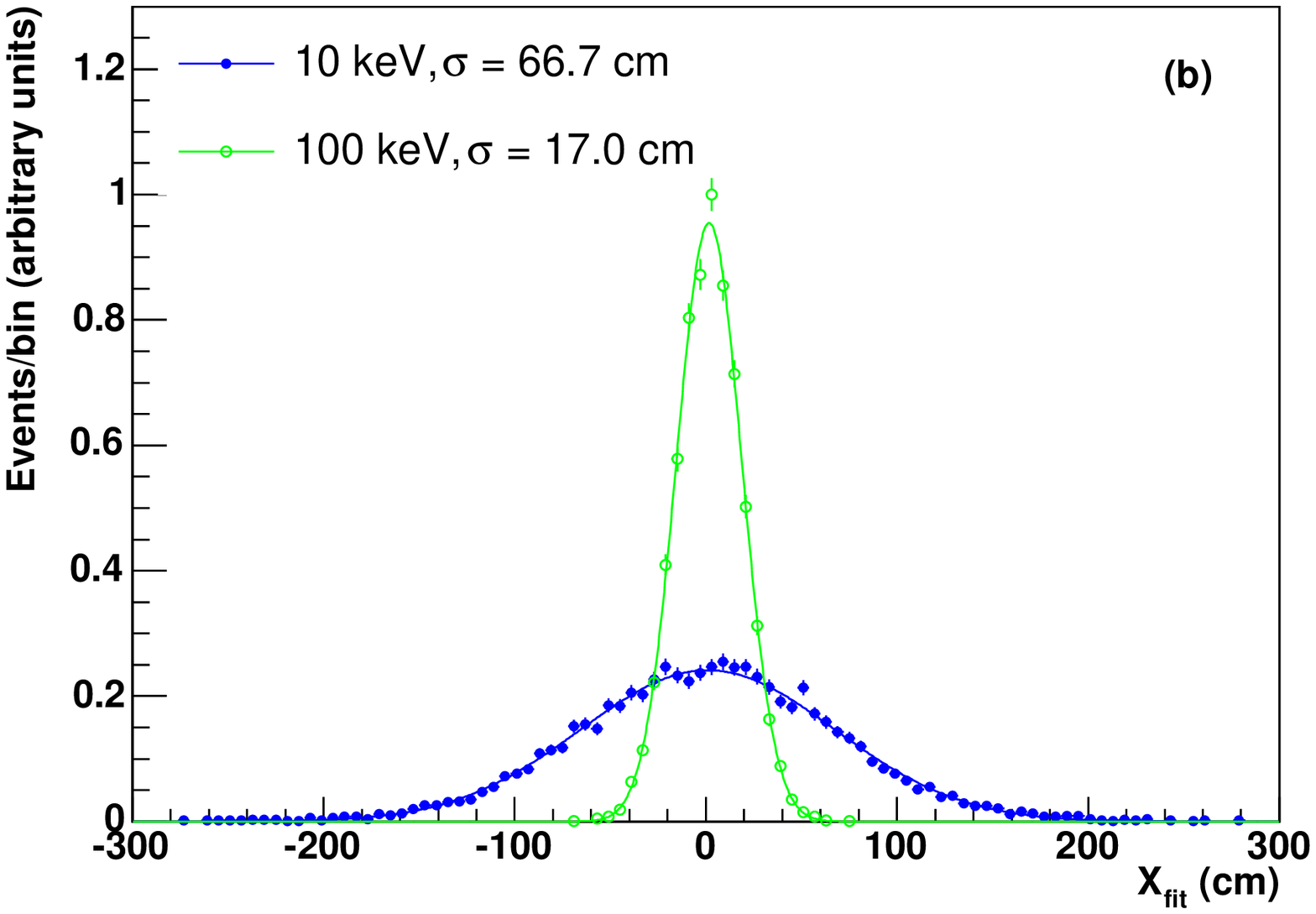}
\caption[Reconstructed energy and position for electrons]{\label{recon_p_and_t}Reconstructed energy (a) and position (b) distributions for simulated electron events.  Position reconstruction shown using spatial information only.}
\end{center}
\end{figure}

\begin{figure}[h]
\begin{center}
\includegraphics[width=\figwidth]{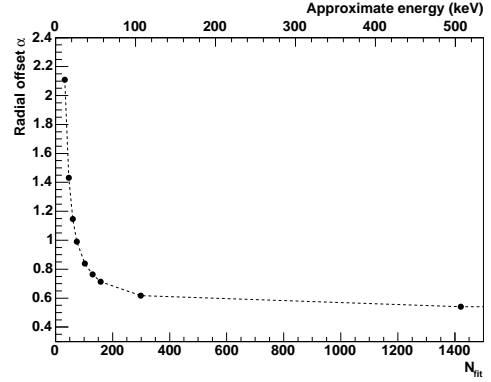}
\caption[Position reconstruction calibration]{\label{pos_recon_cal}Position reconstruction calibration for simulated electron events. Position response is primarily dependent on the number of hit PMTs.  For events with a small number of hits, the dark noise hits pull the fit toward the center.}
\end{center}
\end{figure}

\begin{figure}[h]
\begin{center}
\includegraphics[width=\figwidth]{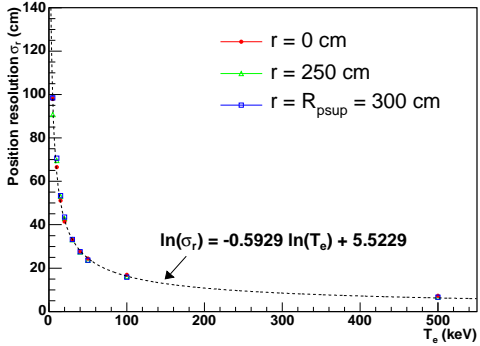}
\caption[Position resolution function]{\label{pos_res}Position resolution function derived from simulated electron events using spatial information.}
\end{center}
\end{figure}

\begin{figure}[h]
\begin{center}
\includegraphics[width=\figwidth]{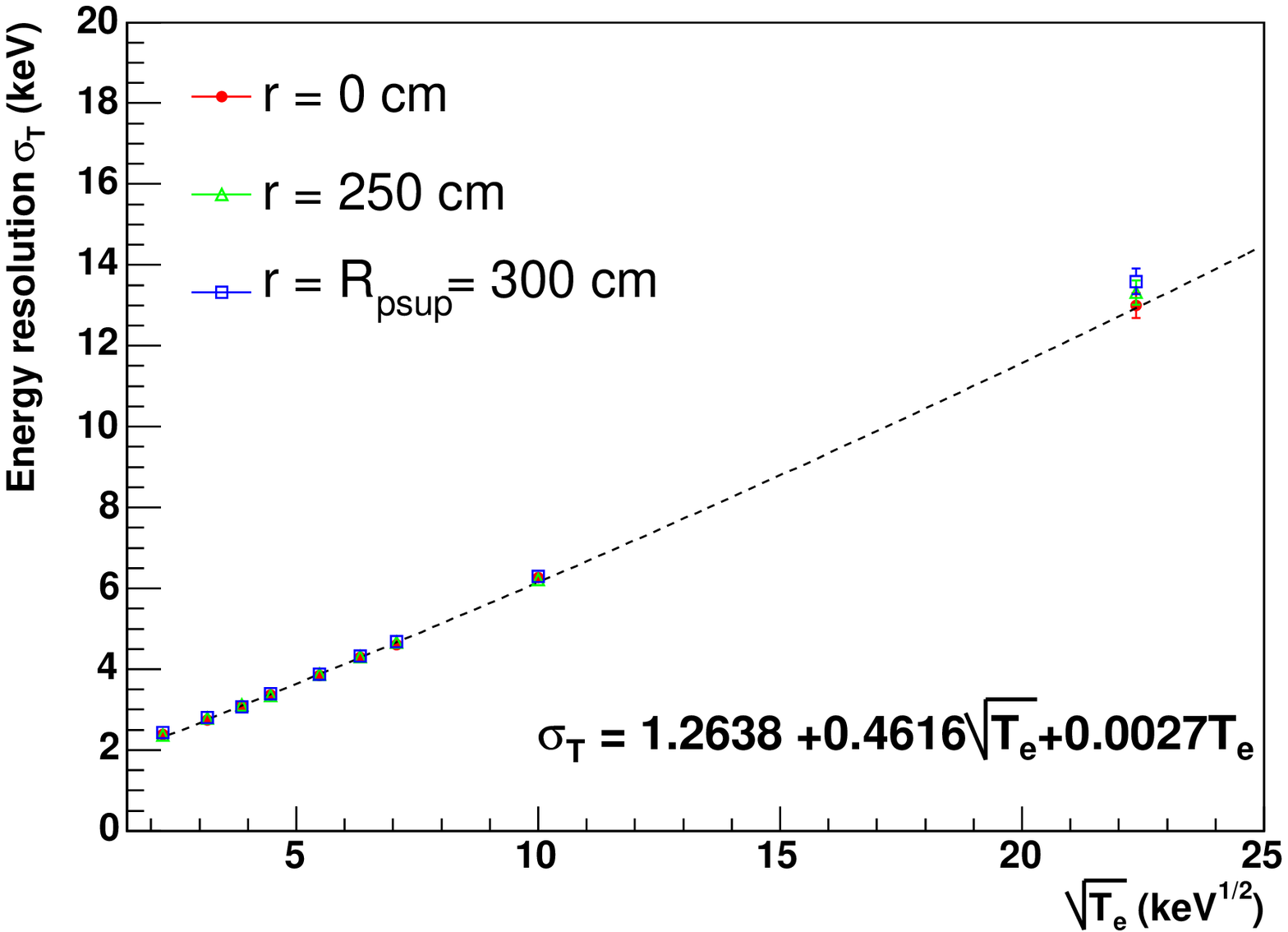}
\caption[Energy resolution function]{\label{teff_res}Energy resolution function derived from simulated electron events.  The energy resolution has little dependence on event position.}
\end{center}
\end{figure}

The reconstructed energy and position distributions (using the spatial algorithm) are shown in Fig.~\ref{recon_p_and_t}.
The reconstruction calibration function ($\alpha$ from equation~\ref{eqn:recon_cal}) is shown in Fig.~\ref{pos_recon_cal}.   The position resolution, derived as a function
of the electron energy, is shown in Fig.~\ref{pos_res}, where it is seen that the position resolution worsens sharply  as the number of noise hits becomes comparable to the
number of scintillation hits.

The energy for each event is estimated by applying the correction
\begin{equation}
T_{\rm eff} = \alpha_{0} + \alpha_{1} \times {\rm NHITs} 
\end{equation}
\noindent where NHITs is the total number of PMT hits, and the energy calibration parameters $\alpha_{0}$ and $\alpha_{1}$ are determined from the Monte-Carlo calculations.\footnote{These parameters will ultimately be calibrated in the physical detector using radioactive sources.}  The energy resolution function, which is
directly related to the total number of hits, is shown in Fig.~\ref{teff_res}.

The distribution of PMT hit times (relative to the trigger time) can be used in the position reconstruction algorithm to improve position resolution.  For a PMT which fired at time 
$t_{pmt}$ for a scintillation event which occurred at time $t_{0}$, we define the time residual
\begin{equation}
t_{res}  =  t_{pmt} - t_{0} - \frac{d_{pmt}}{c} \bar{n} \label{eqn:tres}
\end{equation}
\noindent where
\begin{eqnarray*}
d_{pmt} & = & \mbox{vertex-PMT distance, ~~~} \\
\bar{n} & = & \mbox{average refractive index.}
\end{eqnarray*}

\begin{figure}[h]
\begin{center}
\includegraphics[width=\figwidth]{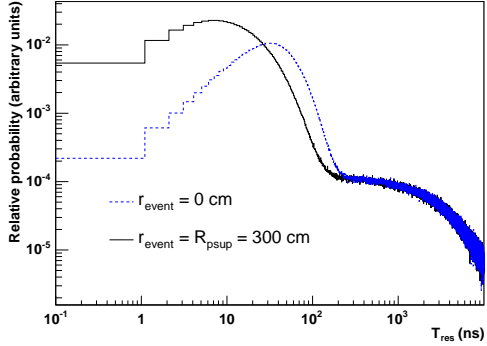}
\caption{\label{tres_two_points}Time residual distribution for simulated 100 keV electron events at r=0 and the PMT radius (r=R$_{\rm psup}$).  The difference in these timing distributions leads to 
the improved position reconstruction.}
\end{center}
\end{figure}

\begin{figure}[h]
\begin{center}
\includegraphics[width=\figwidth]{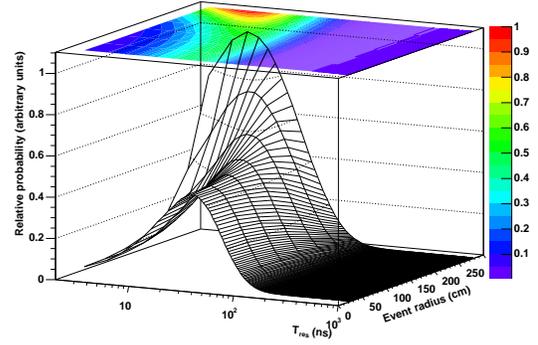}
\caption{\label{2dtrespdf}Two-dimensional time residual pdf derived from Monte-Carlo calculations used in the maximum likelihood position reconstruction algorithm.}
\end{center}
\end{figure}

In the absence of scattering, $t_{res}$ would be the convolution of the scintillation time spectrum with the PMT time resolution.  Scattering will broaden the distribution
and skew it toward later times.  Fig.~\ref{tres_two_points} shows the $t_{res}$ distribution defined by equation~\ref{eqn:tres} for simulated 100 keV electron events at the detector center (r = 0) and near the PMTs (r = R$_{\rm psup}$).  To incorporate timing information
into the reconstruction algorithm, a 2-dimensional probability density function (pdf) in $t_{res}$ versus event radius (shown in Fig.~\ref{2dtrespdf}) has been derived from MC calculations, for which we can calculate the 
probability density for a given event with a given vertex position and time,$P(\vec{x}, t)$.
A likelihood function is defined
\begin{equation}
\mathcal{L}^{'} = \Pi_{i=1}^{N_{sel}}P(\vec{x_f}, t_f;\vec{xi_i}, t_i) \\
\end{equation}
\noindent where $\{\vec{x_{f}}, t_f\}$ are the event position and time, $\{\xi_{i}, t_i\}$ are the set of PMT positions and hit times.  For computational efficiency, we redefine
the likelihood $\mathcal{L}^{'}$ as 
\begin{equation}
\mathcal{L} = -2\sum_{i=1}^{N_{sel}} \log P(\vec{x_f}, t_f; \vec{\xi_i}, t_i) \label{pos_max_like}
\end{equation}

\begin{figure}[h]
\begin{center}
\includegraphics[width=\figwidth]{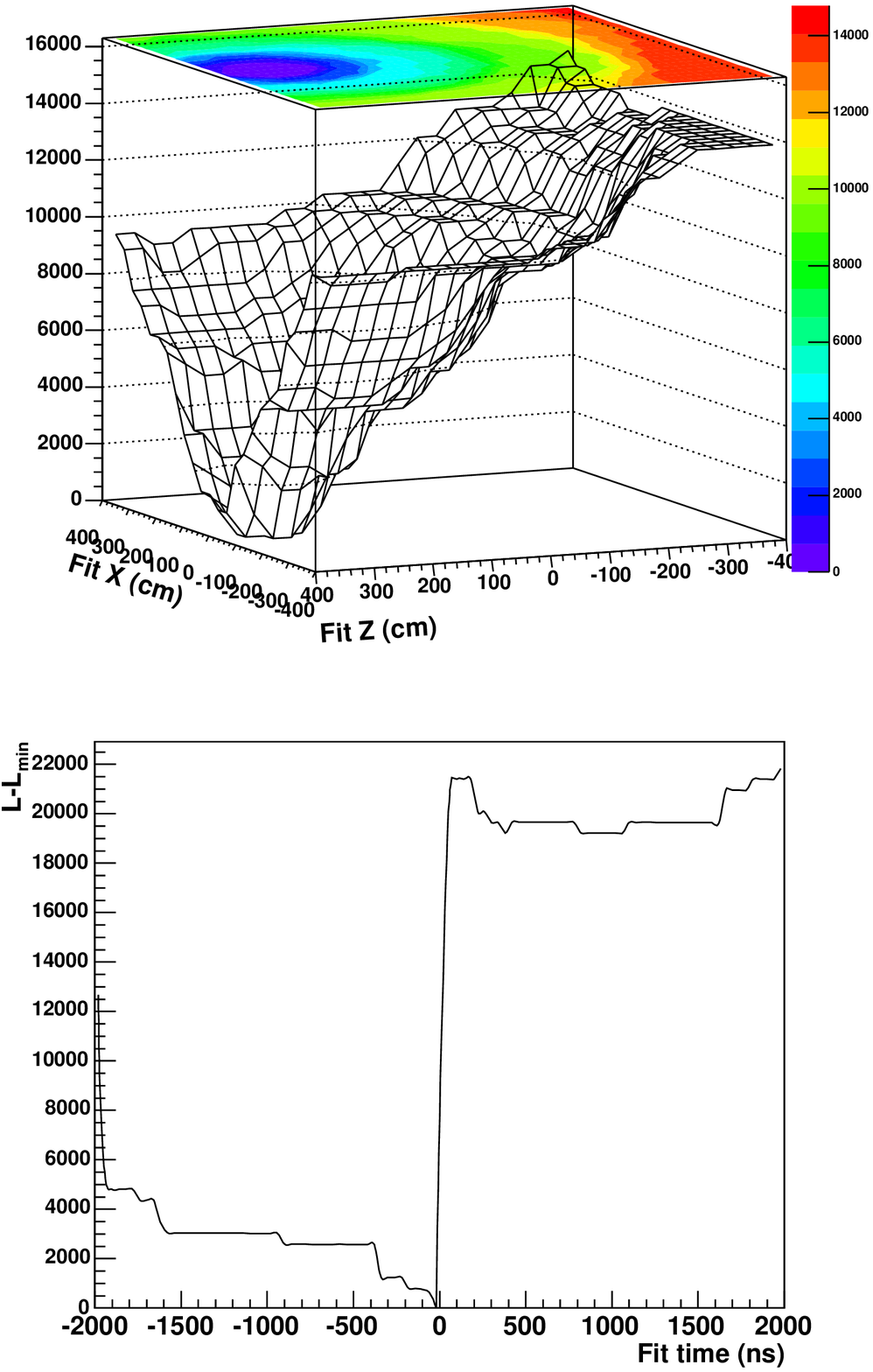}
\caption{\label{sample_like}(top frame) Likelihood surface in x-z plane for simulated 20 keV electron event at position (0, 0, R$_{\rm psup}$), and (bottom frame) likelihood versus event time.  A deep minimum is found near the true event position and time.}
\end{center}
\end{figure}

We can now minimize equation~\ref{pos_max_like} to find the most likely values for $\{\vec{x_f}, t_f\}$.
An x-z slice of the likelihood surface defined by equation~\ref{pos_max_like} is shown in Fig.~\ref{sample_like}(a) for a simulated
20 keV electron event at position (0, 0, R$_{\rm psup}$), and the likelihood versus event time is shown in Fig.~\ref{sample_like}(b),  where a deep minimum is seen near the event position and time.  The information in the spatial
hit pattern of the PMTs can be added to the likelihood function by including to equation~\ref{pos_max_like} the position response
using only the hit information:
\begin{equation}
\mathcal{L}^{''} = [\Pi_{i=1}^{N_{sel}} P(\vec{x_f}, t_f; \vec{\xi_i}, t_i)] \times P_{s}(\vec{x_f}, \vec{\xi_i} ) \label{eqn:full_like}
\end{equation}
\noindent where $P_{s}(\vec{x_f}, \vec{\xi_i} )$ is the position response function shown in Fig.~\ref{pos_res}.

\begin{figure}[h]
\begin{center}
\includegraphics[width=\figwidth]{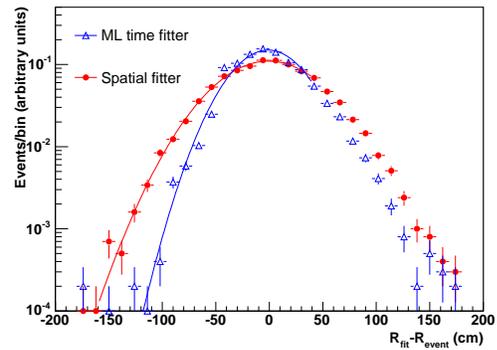}
\caption{\label{compare_fit_alg}Reconstructed event positions for spatial and ML time fitters for 20 keV electron events at r = R$_{\rm psup}$.  }
\end{center}
\end{figure}

\begin{figure}[h]
\begin{center}
\includegraphics[width=\figwidth]{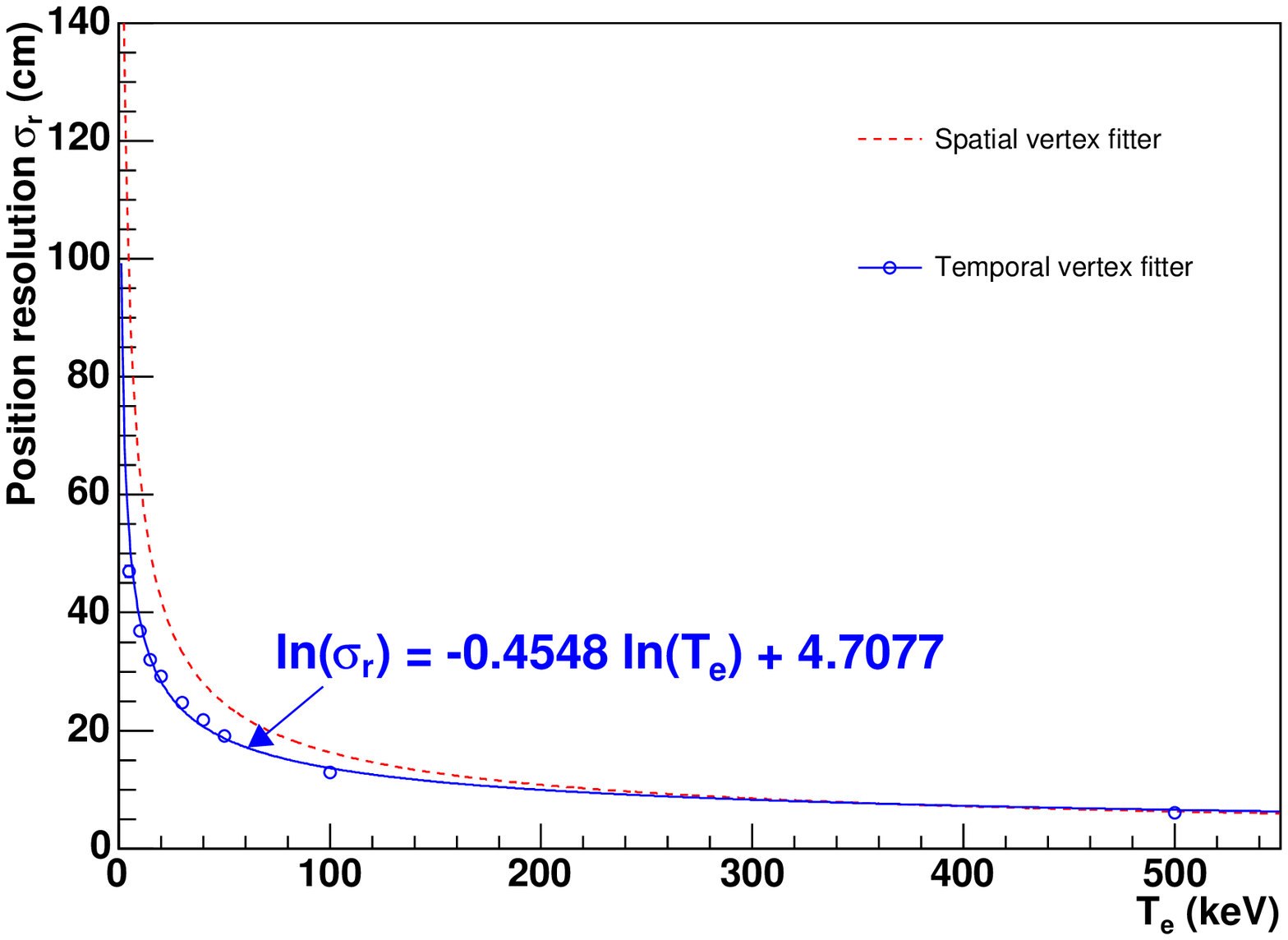}
\caption{\label{compare_pos_res_fns}Position resolution with spatial and temporal reconstruction algorithms.  Including time information improves position resolution by a factor of approximately two for low energy events.}
\end{center}
\end{figure}

Fig.~\ref{compare_fit_alg} shows the reconstructed position distributions using equation~\ref{eqn:full_like} and equation~\ref{r_est} (spatial only) for 20 keV electron events
at r = R$_{\rm psup}$.  The resolution for low-energy events improves by a factor of approximately two when the time information is included.  This is a significant
improvement since it implies the same background reduction as doubling the detector size, 
assuming 20 keV background at the PMT radius.  The derived position resolution functions versus kinetic energy for both methods are shown in Fig.~\ref{compare_pos_res_fns}.

The response functions shown in Figures~\ref{teff_res} and~\ref{compare_pos_res_fns} will be used in the following sections to predict the energy and position distributions for signal
and background events in the detector.  The calibration requirements for energy and position reconstruction are discussed in section~\ref{section_calibration}.

\section{Solar neutrino sensitivity}
Solar neutrinos are detected by their elastic scattering from electrons in the target material.  The recoiling electrons will generate scintillation light.
The predicted differential electron recoil spectrum from elastic scattering of solar neutrinos is 
\begin{equation}\begin{split}
\frac{dN}{dT_{e}}& = n  \int_{0}^{\infty}  [ \lambda_{e}(T^{'})\frac{d\sigma_{e}}{dT^{'}}(T^{'},T_{e}) \\
&+\lambda_{\mu\tau}(T^{'}) \frac{d\sigma_{\mu\tau}}{dT^{'}}(T^{'},T_{e}) ] dT^{'} 
\end{split}\end{equation}
\noindent with
\begin{eqnarray*}
 n & = & \mbox{number of electrons in target} \\
\lambda_{x} & = & \mbox{$e$ or $\mu,\tau$ component of $\nu$ flux} \\
\frac{d\sigma_{x}}{dT^{'}} & = & \nu-e^{-}\mbox{cross section} 
\end{eqnarray*}

\begin{figure}[h]
\begin{center}
\includegraphics[width=\figwidth]{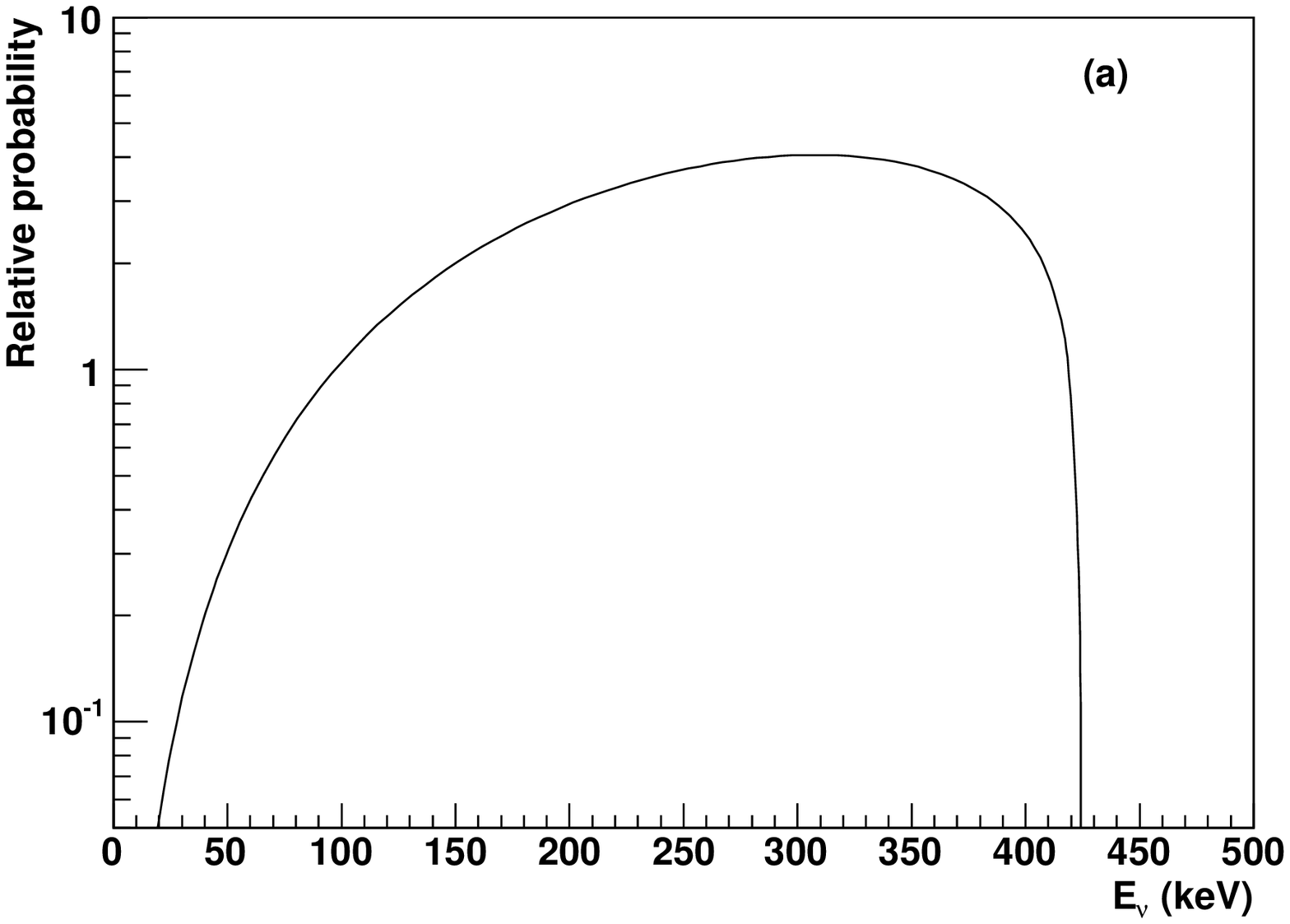}
\includegraphics[width=\figwidth]{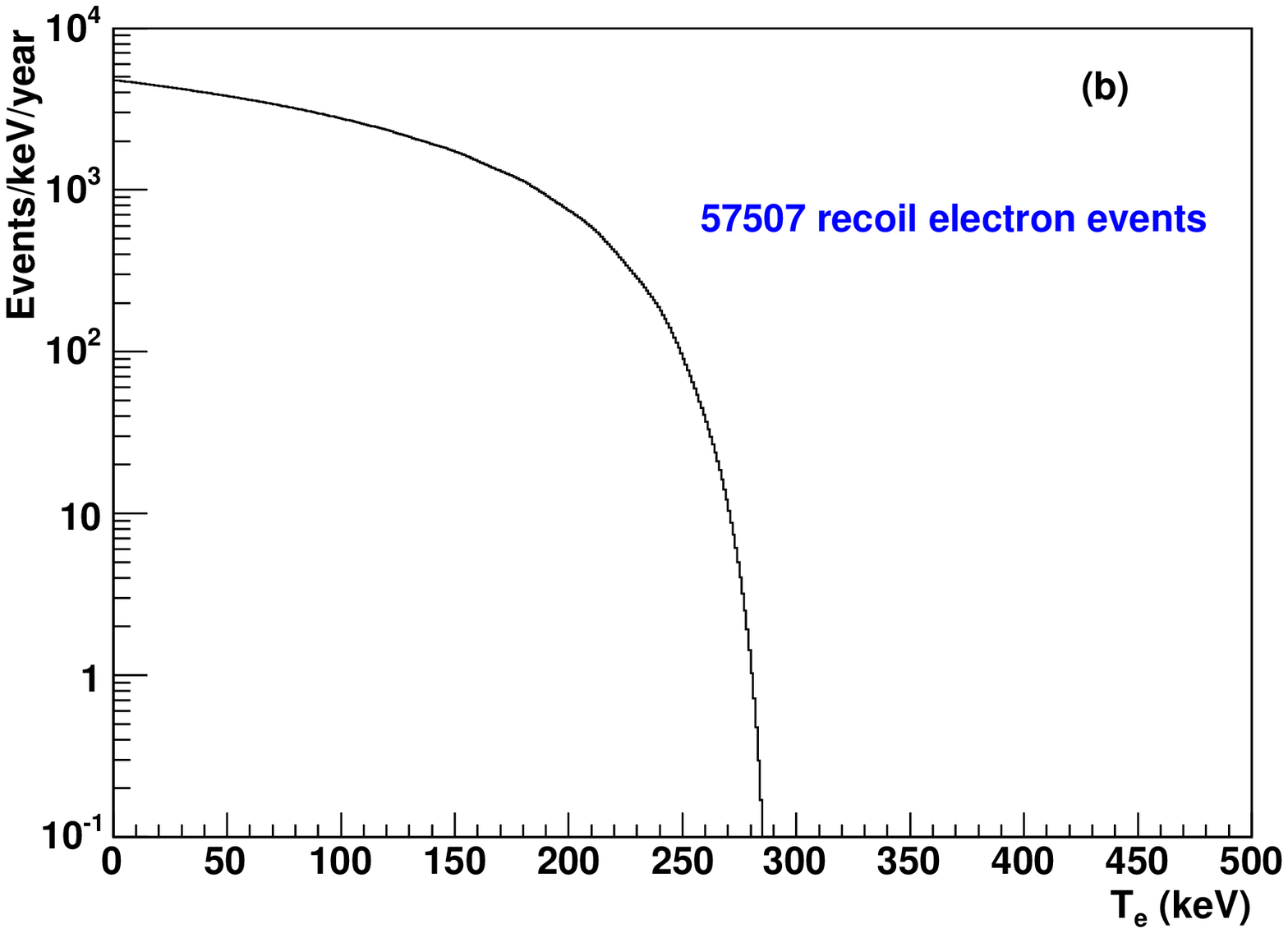}
\includegraphics[width=\figwidth]{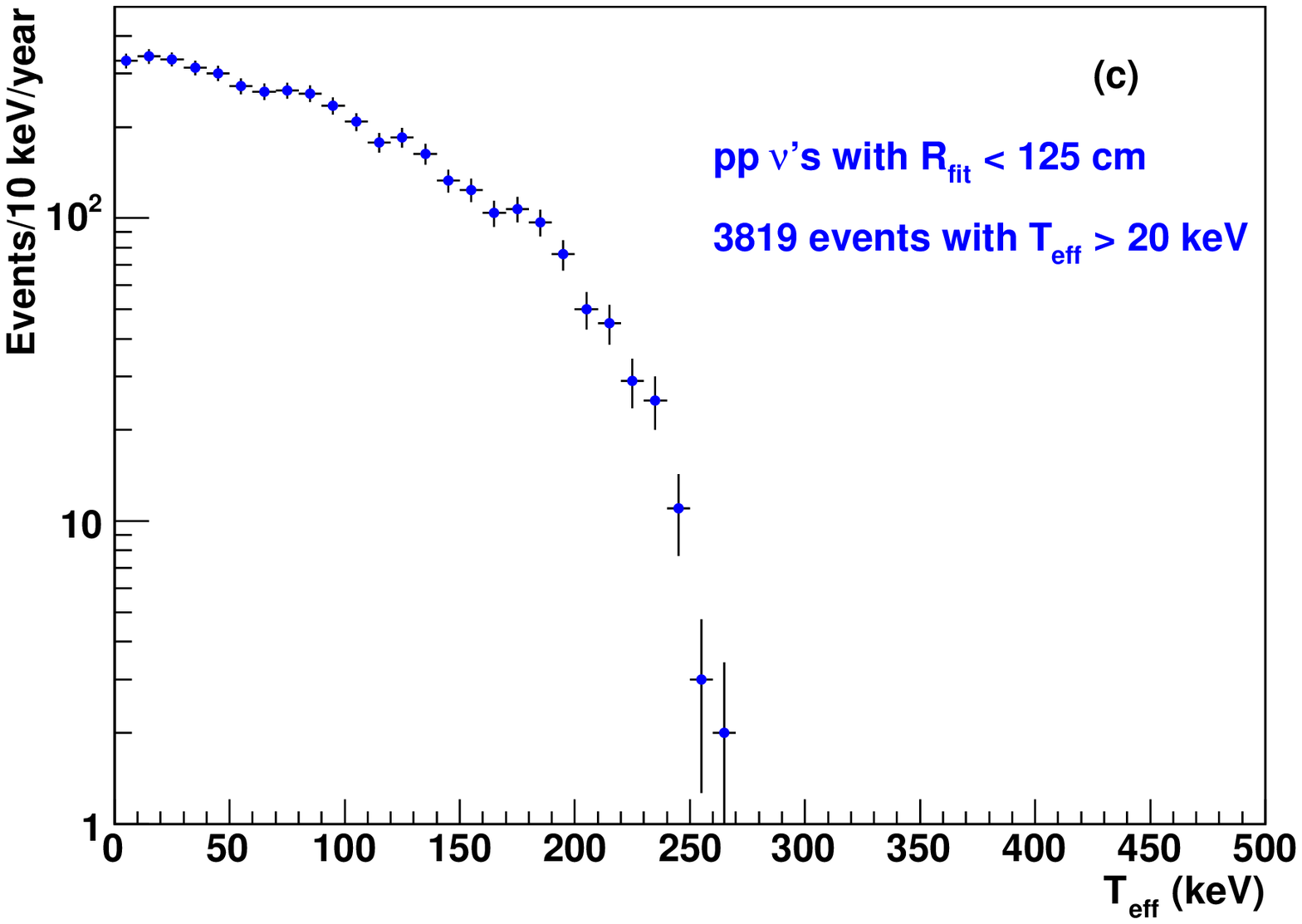}
\caption[Predicted solar neutrino spectrum]{\label{nu_spec}(a) Solar $pp$ neutrino energy spectrum predicted by Bahcall {\it{et. al.}}~\cite{bp2000}. (b) Predicted electron recoil energy distribution in liquid neon. (c) Predicted observed energy distribution with a 125 cm fiducial volume cut (approx. 10 tonnes).   Figures (b) and (c) are normalized to 1 live year.}
\end{center}
\end{figure}

\noindent The neutrino-electron elastic scattering cross-section is taken from~\cite{bahcall_book}.  The $e$ and \{$\mu,\tau$\} components of the solar neutrinos can be estimated by using the
standard solar model predicted spectra and rates~\cite{bp2000}, and the solar neutrino mixing parameters from a global analysis of the solar and reactor neutrino experiments~\cite{salt_prl}.
With the parameters from Table~\ref{solar_pars}, we can calculate the predicted solar neutrino spectrum as
\begin{eqnarray}
\lambda_{e}(T) & = & \lambda(T)P_{ee}(T) \\
\lambda_{\mu\tau}(T) & = & \lambda(T)(1-P_{ee}(T)) 
\end{eqnarray}
\noindent where $\lambda(T)$ is the undistorted solar neutrino spectrum, and $P_{ee}$ is the electron neutrino survival probability at Earth calculated with the nominal solar mixing parameters.\footnote{The survival probability was calculated with analysis code provided by the SNO collaboration.}
Fig.~\ref{nu_spec}(a) shows the predicted $pp$ solar neutrino energy spectrum.  Fig.~\ref{nu_spec}(b) shows the predicted differential electron recoil energy spectrum for a 3 meter radius target of liquid neon exposed to $pp$ neutrinos for 1 year, and figure~\ref{nu_spec}(c) shows the predicted
spectrum measured with the detector, for a fiducial volume cut of 125 cm, representing approximately 10 fiducial tonnes of target.  The total number of events within these cuts is 3819, which leads to a statistical uncertainty on the neutrino flux of approximately 1.6\%.

\begin{table}[htb]
\begin{center}
\caption[Solar neutrino parameters]{\label{solar_pars}Solar neutrino absolute flux and mixing parameters.}
\begin{tabular}{ll} \hline
Parameter & value \\ \hline
$pp$ flux & $5.95 \times 10^{10}$ cm$^{-2}$s$^{-1}$ \\
$\Delta m^{2}$ & $7.1^{+1.2}_{-0.6} \times 10^{-5}$ eV$^{2}$ \\
$\theta$ & $32.5^{+2.4}_{-2.3}$ degrees \\ \hline 
\end{tabular}
\end{center}
\end{table}

\clearpage
\section{Backgrounds}
\subsection{External backgrounds}
Purification of the neon is expected to reduce background contamination from internal sources of radioactivity significantly~\cite{clean_paper}.  The PMTs and associated hardware will contain the largest amount of radioactive contamination near the inner detector volume.  To reduce these backgrounds we must design an active shielding region of liquid neon, and apply
position reconstruction cuts to define a fiducial signal region with acceptable background levels.
With the nominal impurities in the PMTs, the number of decays from the uranium and thorium chains, and from ${}^{40}$K is large, shown in Table~\ref{pmt_decays}.
In principle we would like to simulate all of the activity from these sources of background, and apply event reconstruction and fiducial volume cuts to estimate
the residual background contamination in the signal region.  This is difficult in practice given finite computational resources, since the total number of simulated
events would be large.

\begin{table*}[h]
\begin{center}
\caption[Decays in PMTs]{\label{pmt_decays}Decay rates in D737KB PMTs.}
\begin{tabular}{llll} \hline
Chain & Specific activity & Total mass in PMTs & Decays/year \\
      & (Bq/g)            & (g)                &             \\
U     & $178.9\times10^{3}$                  &   0.0553                 & $3.122\times10^{11}$             \\
Th    & $39.14\times10^{3}$                  &  0.0553                  & $6.83\times10^{10}$            \\ 
K     & $30.30$                  & 110.52                   & $1.057\times10^{11}$ \\ \hline
\end{tabular}
\end{center}
\end{table*}

\begin{figure}[h]
\begin{center}
\includegraphics[width=\figwidth]{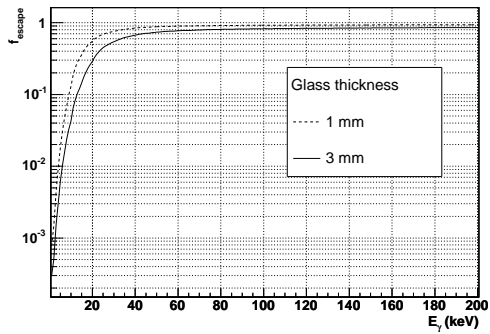}
\caption{\label{fescape}Photon escape fraction from PMT glass versus photon energy, for 1mm and 3mm thick glass. Low-energy x rays are highly attenuated by the glass.}
\end{center}
\end{figure}

\noindent To decrease computational requirements, the detector response to x-ray and $\gamma$-ray photons from the uranium and thorium chains and from potassium decays in the PMTs is estimated by
using the MC-calculated response functions in energy and radius from section~2:
\begin{equation}\begin{split}
\frac{d^{2}N}{dTdr} &= \int_{T^{'}} \int_{r^{'}}  \gamma(T^{'},r^{'}) R(T, T^{'}, r, r^{'}) \\
& f_{escape}(T^{'})\Omega dT^{'}dr^{'} \label{gamma2ddist}
\end{split}\end{equation}
\noindent where $\gamma(T^{'},r^{'})$ is the distribution of energy deposition (in kinetic energy $T^{'}$ and radius $r^{'}$) due to 
photons from the PMTs, and $R(T, T^{'}, r, r^{'})$ is the detector response function. $f_{escape}$ is a correction factor for photon absorption in the PMT glass, shown in Fig.~\ref{fescape}, 
and $\Omega$ is a solid angle correction to account for x-ray and $\gamma$-ray coincidence (approximately  0.5 for x rays and 1 for $\gamma$ rays).

\begin{figure}[h]
\begin{center}
\includegraphics[width=\figwidth]{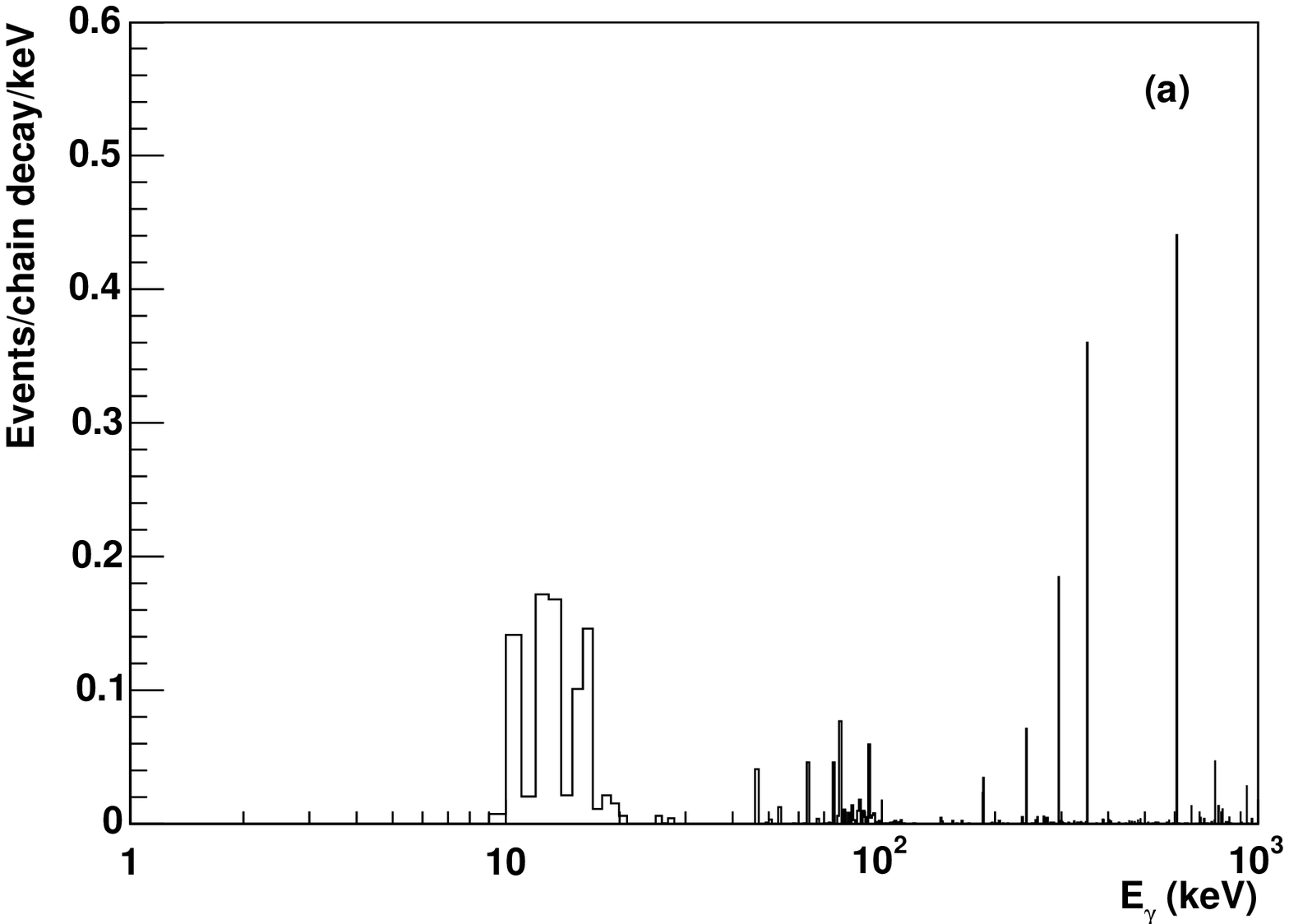}
\includegraphics[width=\figwidth]{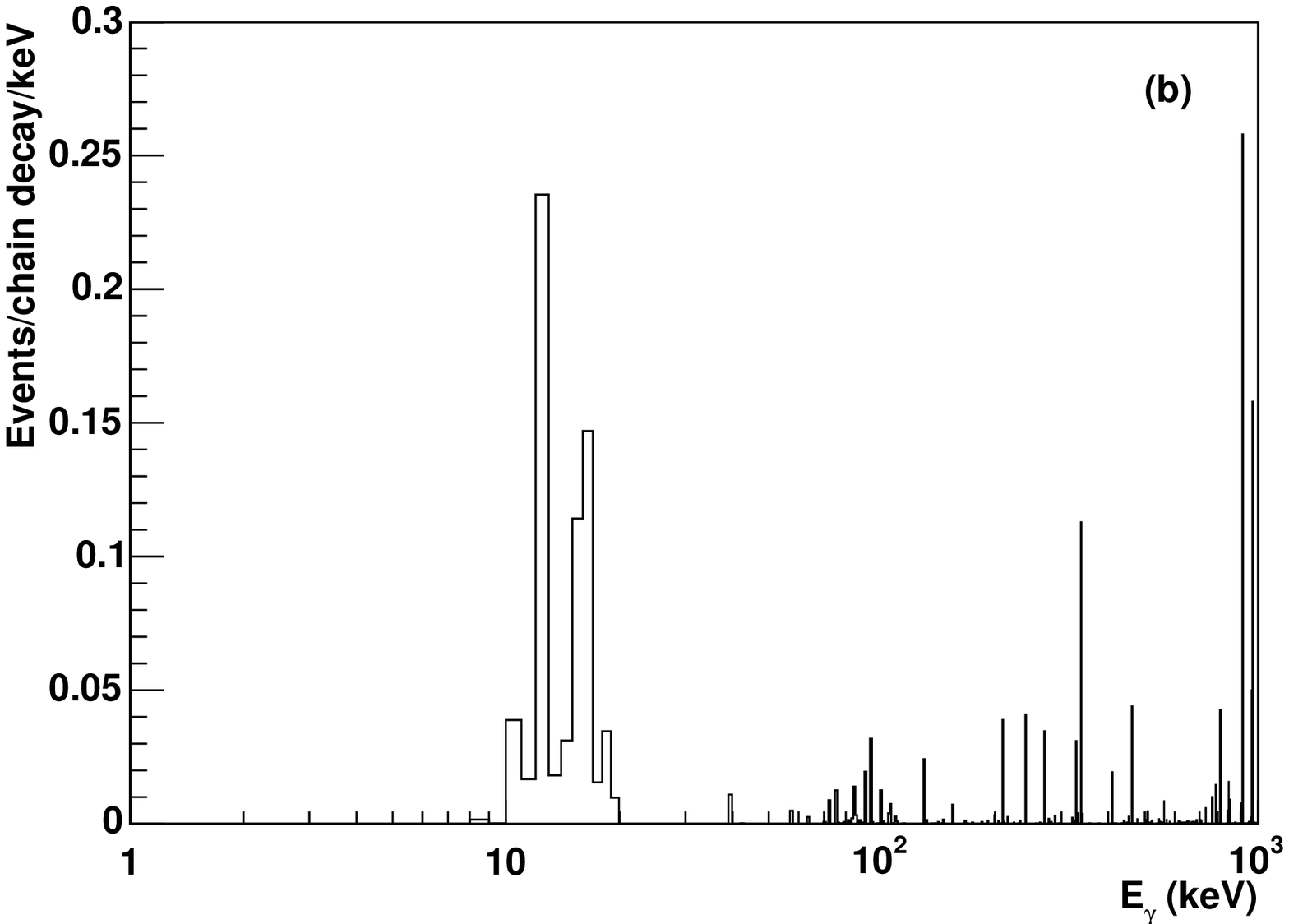}
\includegraphics[width=\figwidth]{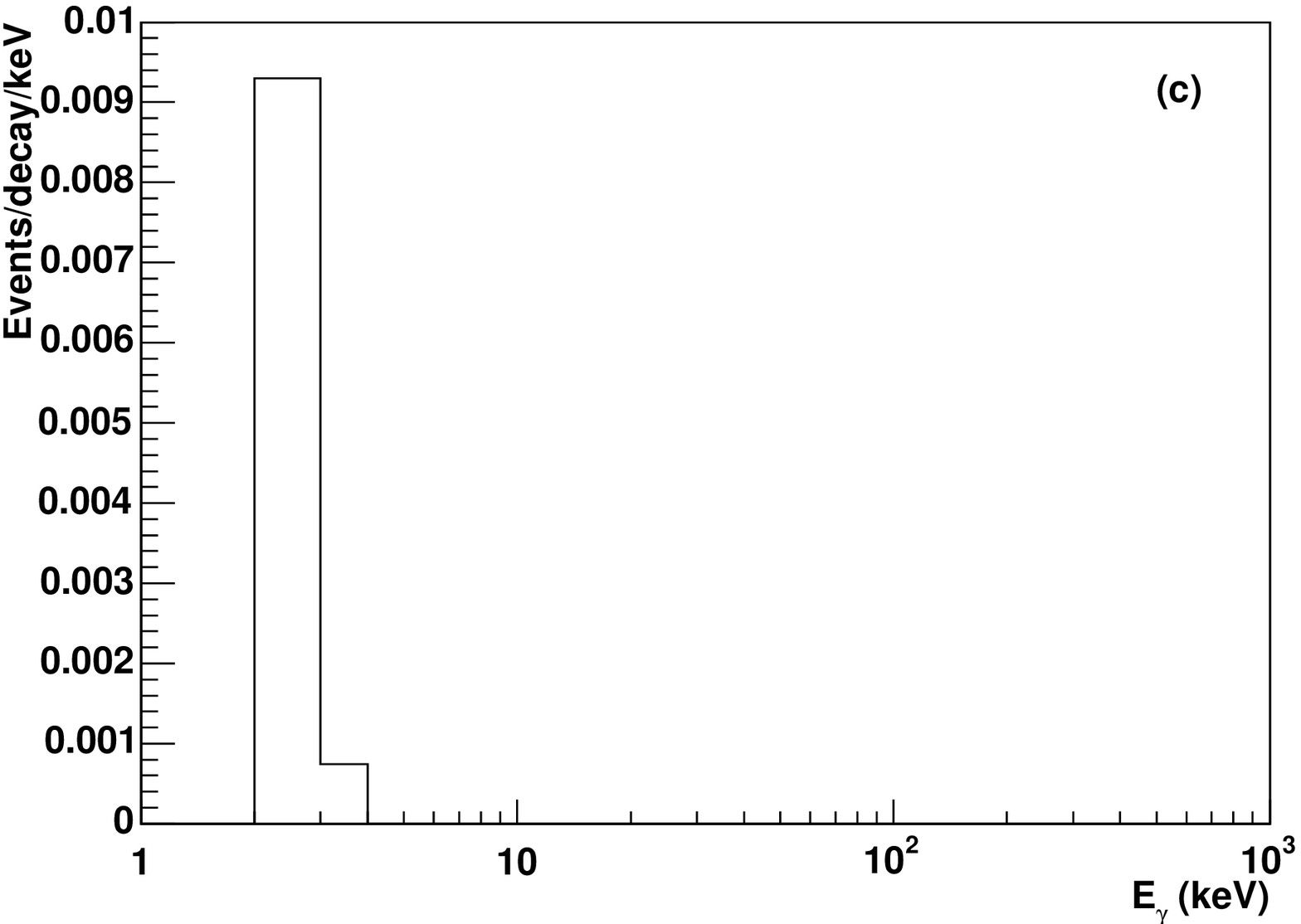}
\caption[$\gamma$ and x-rays from uranium and thorium chains]{\label{uthchain_gammas}$\gamma$ and x-rays from uranium (a) and thorium (b) chain decays, and from potassium (c).  Shown are the number of events per keV per chain decay [in (a) and (b)], and per decay [in (c)]. Approximately 5$\%$ of each chain decay results in a photon in the range 0-50 keV.}
\end{center}
\end{figure}

\begin{figure}[h]
\begin{center}
\includegraphics[width=\figwidth]{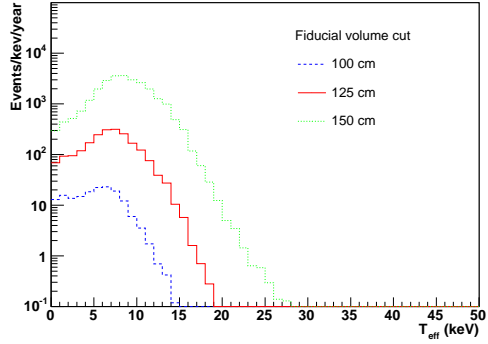}
\caption[External backgrounds from U and Th]{\label{pmt_back_fid}External background from contamination with D737KB PMTs at 300 cm, for different fiducial volume cuts.}
\end{center}
\end{figure}

\begin{figure}[htb]
\begin{center}
\includegraphics[width=\figwidth]{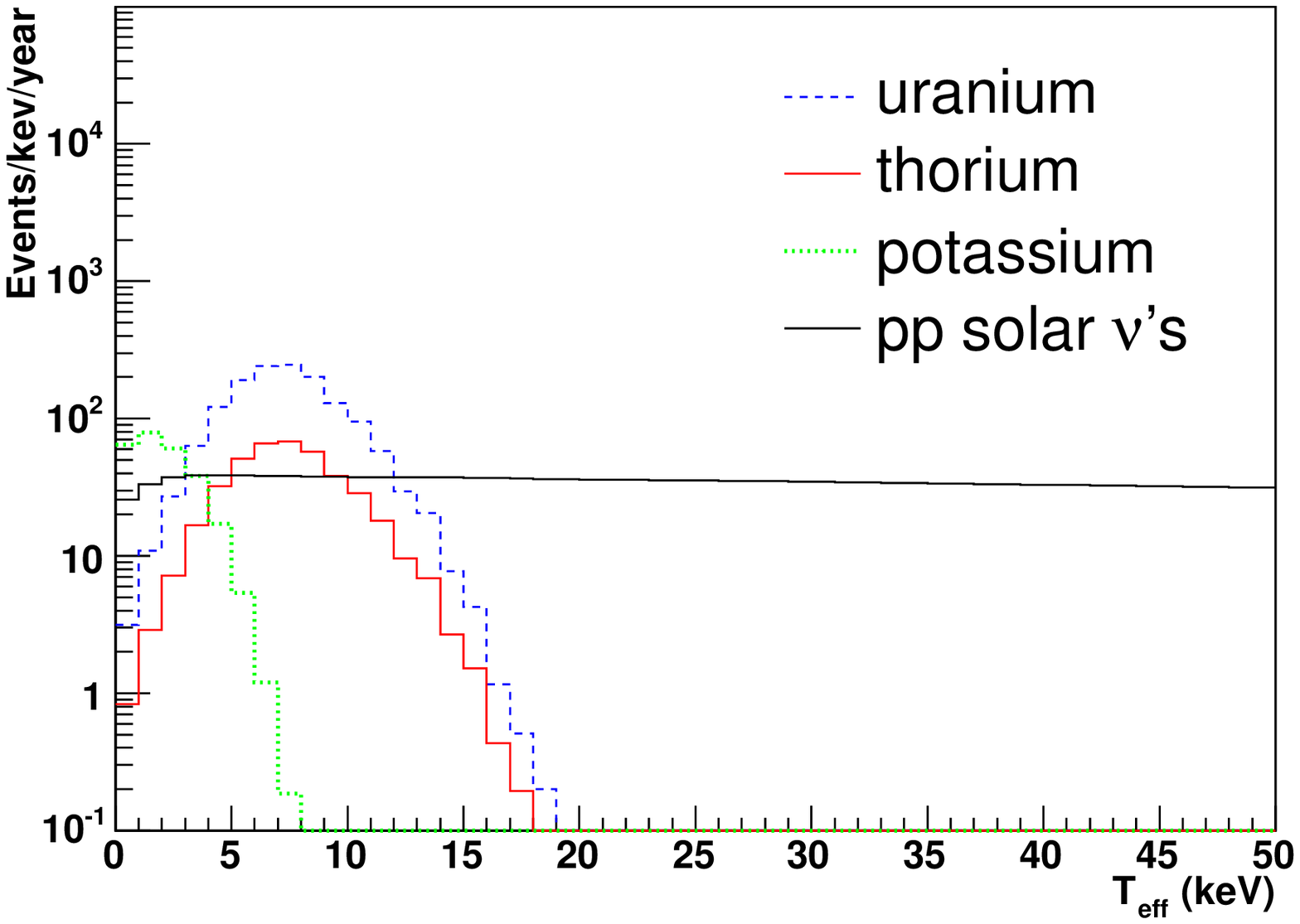}
\caption{\label{back_comp}Backgrounds from uranium, thorium, and potassium for a PMT sphere with radius 300 cm, and a 125 cm fiducial volume.}
\end{center}
\end{figure}

 We further assume
\begin{equation}
R(T, T^{'}, r, r^{'})  =  R(T,T^{'}) \times R(r,r^{'}, T^{'})
\end{equation}
and use the response functions shown in Figs.~\ref{compare_pos_res_fns}~and~\ref{teff_res}.\footnote{Note the subtle difference between $R(r,r^{'},T)$ and $R(r,r^{'},T^{'})$.}
In this scheme we need only the position and energy response functions derived, and a model for the $\gamma$-ray distribution at the PMTs due to radioactive contamination in order to estimate
residual backgrounds.
The distribution $\gamma(T^{'},r^{'}$) is calculated with Monte-Carlo by generating the photon energy distribution from the uranium and thorium chains, and from potassium decays (including all x and $\gamma$ rays listed in the Table of radioactive isotopes~\cite{toi}, shown in Fig.~\ref{uthchain_gammas}) and recording the resulting radial and energy distributions.  Low-energy x-rays are most problematic even though they are strongly attenuated by the PMT glass and the liquid neon, since the poor position resolution
at low energies allows these events to misreconstruct within the fiducial volume.  
We can integrate equation~\ref{gamma2ddist} between $r=0$ and $r=r_{0}$ to estimate the effective energy spectrum due to 
decays in the PMTs for a fiducial volume cut, $r_{0}$.  Fig.~\ref{pmt_back_fid} shows these for several fiducial volume cuts. 
 Fig.~\ref{back_comp} shows the PMT backgrounds
for each contaminant, and the intersection with the $pp$ energy spectrum which defines the analysis threshold.  The achievable background reduction for the commercially available PMTs would allow a low enough threshold (approximately 13 keV) for
a feasible $pp$ neutrino and dark matter experiment. 

\subsection{Internal backgrounds}
\begin{figure}[h]
\begin{center}
\includegraphics[width=\figwidth]{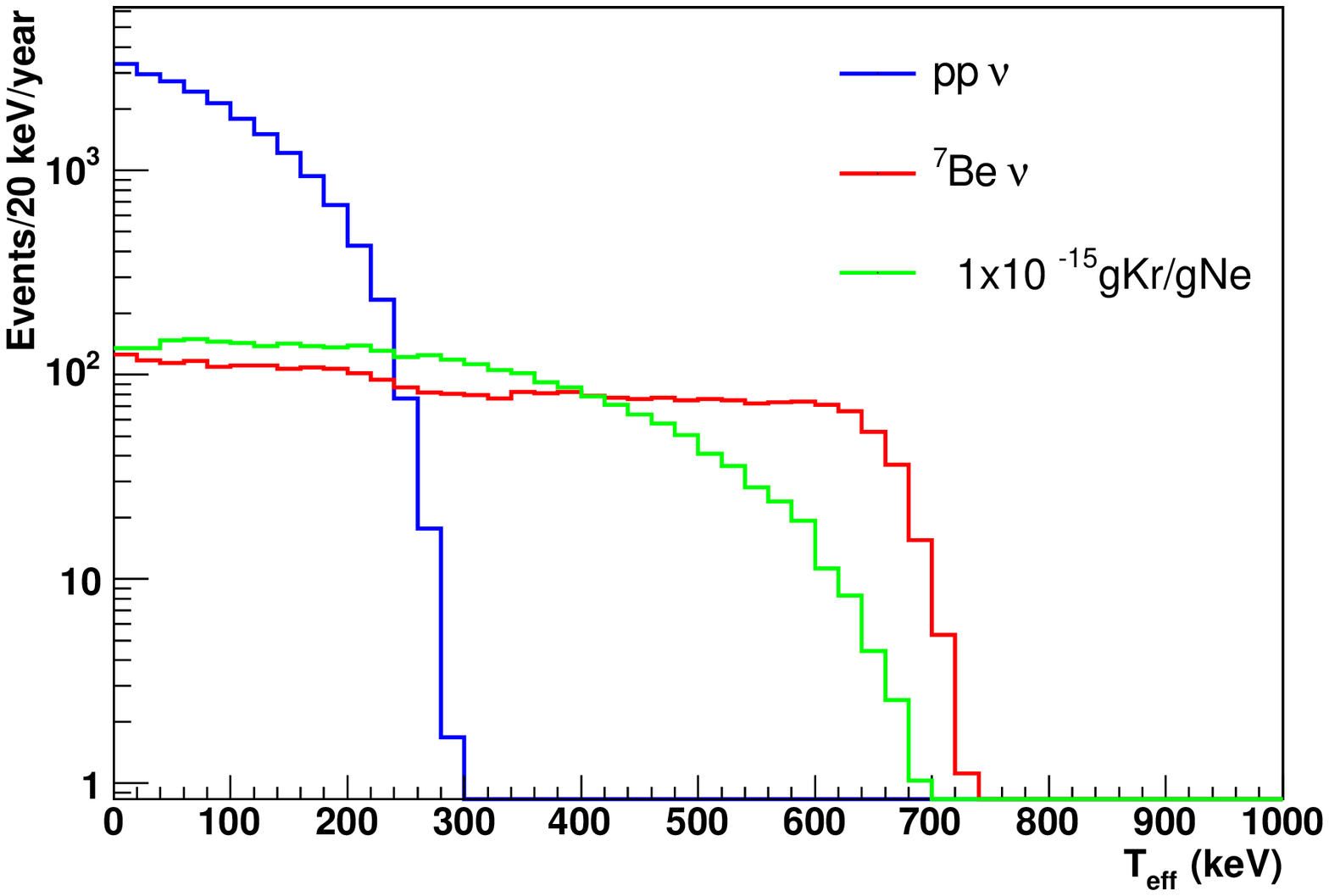}
\includegraphics[width=\figwidth]{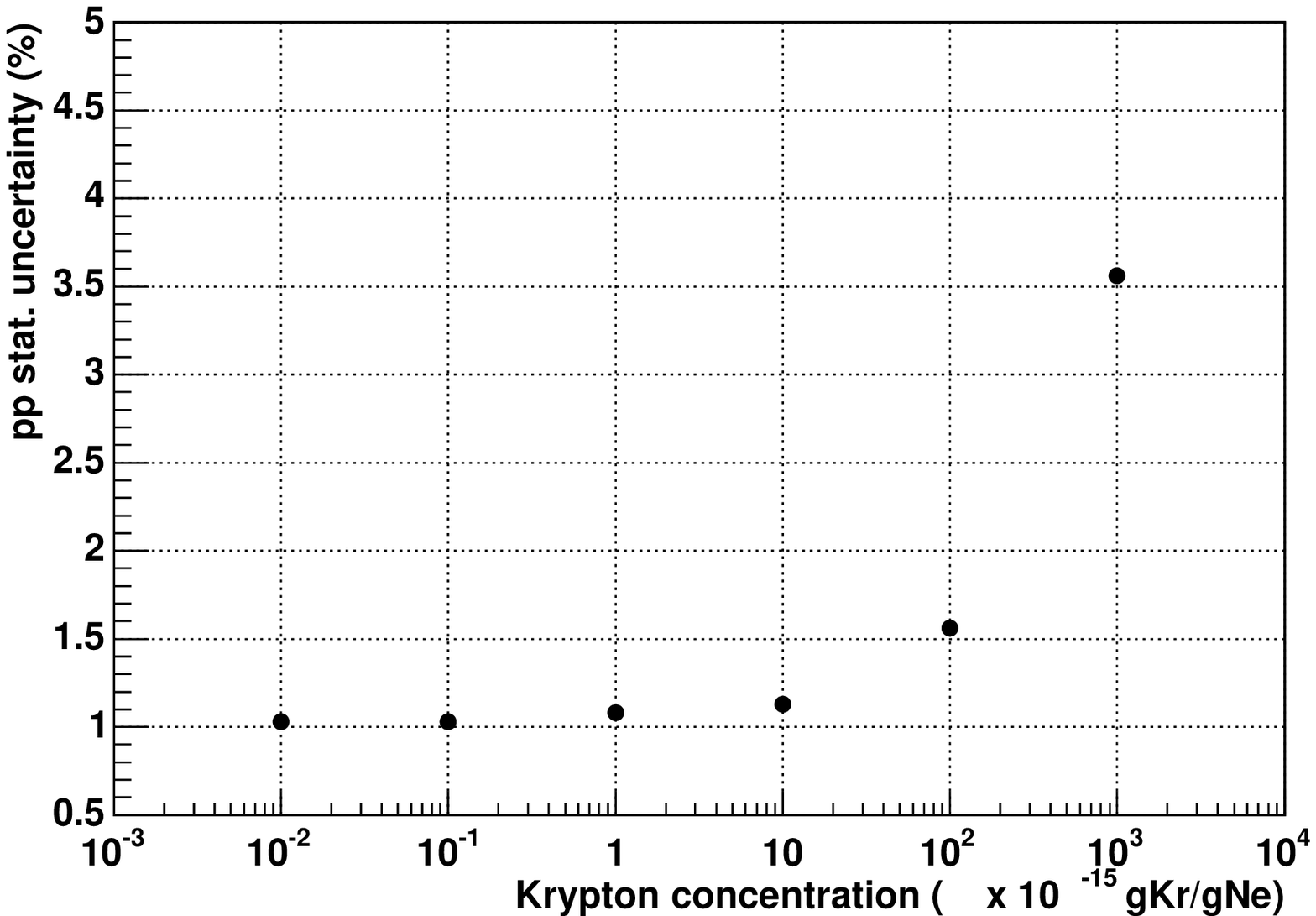}
\caption{\label{krinsitu}{\it{In-situ}} determination of ${}^{85}$Kr background.  (top)  T$_{\rm eff}$ spectra predicted for $pp$ and ${}^{7}$Be neutrinos, and for ${}^{85}$Kr.
(bottom) Uncertainty on $pp$ flux from {\it{in-situ}} determination versus Kr concentration.}
\end{center}
\end{figure}

The dominant internal source of background for the $pp$ neutrino measurement is expected to be ${}^{85}$Kr since it has a relatively short
half-life ($\approx$ 11 years), is a $\beta$ emitter with energies in the same range (Q-value = 687 keV) as the $pp$ neutrinos, and is present in the atmosphere.  Several
other naturally-occurring radioactive contaminants, expected to be less troublesome than ${}^{85}$Kr, will need to be removed from the neon in order to achieve acceptable background levels for neutrino detection, most notably uranium and thorium. 
To estimate the requirements for internal background contamination, numbers of events from the uranium chain, thorium chain, and from ${}^{85}$Kr were estimated, and target concentrations were set which would lead to less than $1\%$ background to the solar $pp$ measurement between 20-500 keV (the region of interest for $pp$ neutrinos).  These target concentrations are shown in Table~\ref{internal_bg_limits}. 
 The required concentrations for uranium and thorium, although quite low, have been achieved in large quantities
of organic liquid scintillator~\cite{borexino}.  Due to the different binding energy of neon on carbon relative to those of most other impurities~\cite{vivaldi}, cryogenic purification with cold traps is expected to be
quite effective.  The requirements for ${}^{85}$Kr are severe and {\it{ex-situ}} assays at this level will present unprecedented
challenges.  The spectral shape of $\beta$ decays from ${}^{85}$Kr, shown in Figure~\ref{krinsitu}(a) can be used to extract this activity
{\it{in-situ}} from the PMT data.  Figure~\ref{krinsitu}(b) shows the inferred statistical uncertainty on the integral $pp$ neutrino flux
from a maximum likelihood (ML) separation of ${}^{85}$Kr decays, ${}^{7}$Be neutrino, and $pp$ neutrino interactions.  It is found that a concentration
of $10^{-15}$ of ${}^{85}$Kr can be extracted {\it{in-situ}} without significantly increasing the $pp$ uncertainty, although the correlation with 
the ${}^{7}$Be signal will make separation of ${}^{85}$Kr from ${}^{7}$Be neutrino interactions difficult.

\begin{table*}[htb]
\begin{center}
\caption[Internal radioactivity requirements]{\label{internal_bg_limits}Internal radioactivity requirements.  The target mass and concentrations are for a background of 
less than $1\%$ of the $pp$ solar neutrino rate with an energy threshold of 20 keV.}
\begin{tabular}{llllll} \hline
Nuclide & Activity & Isotopic  & Specific & Target  & Target  \\
        & source   & abundance & activity & mass        & Conc. \\ 
        &          &           & (Bq/g)   & (ng)       &  (g/g) \\ \hline
Kr      & $^{85}$Kr                  & 1.5e-11          & $1.55\times10^{6}$     & $3.7\times10^{-4}$& 1 $\times 10^{-15}$ \\
Th      & ${}^{232}$Th chain           & 1.00             & $39.14\times10^{3}$    & 3.1               & 2.2 $\times 10^{-17}$ \\
U       & $^{235}$U, $^{238}$U chains& 0.993, 0.007 & $178.9\times10^{3}$    & 8.5               & 6.2 $\times 10^{-17}$ \\ 
\end{tabular}
\end{center}
\end{table*}

\subsection[Cosmogenic backgrounds]{Cosmogenics and neutron backgrounds}

The primary backgrounds for dark matter and solar neutrino experiments related to cosmic ray muons are fast neutrons produced in materials
surrounding the detector and long-lived radioactivity created in the detector itself from muon spallation.  Mei~\cite{mei_paper} has estimated
the size of these backgrounds for a liquid neon experiment and concludes that  they can be made negligible
by locating the detector at sufficient depth (greater than 5000 meters water equivalent) to reduce the incoming muon flux.

The large spherical volume of neon will also act as a veto for neutrons from sources external to the detector which penetrate a water shielding region.  Neutrons will
thermalize and be captured in short distances in liquid neon.  A fiducial volume cut can thus be used to remove
external neutrons, and the coincidence signal between the scintillation photons from neutron-neon recoils and the subsequent
capture $\gamma$ ray can further reduce this background.  In addition the time distribution 
of a neutron event will be distinguishable from a               WIMP-nucleus scatter, since a neutron in liquid neon will scatter several times, while a WIMP will scatter only once.  The neutron time distribution is thus broader than the corresponding distribution
for a WIMP-induced recoil.  In this study the detector is assumed to be at a sufficient depth and have adequate shielding so that these backgrounds are negligible.

\section{Dark Matter Sensitivity}
In the approach taken here, the ultimate sensitivity to dark matter is determined by the limit on the total rate of detected dark matter interactions.  This rate is a function of the WIMP-nucleon
cross-section, the fiducial target mass, the analysis energy threshold, and the number of irreducible background events.  We present here the predicted rates and recoil energy spectra for WIMP-nucleon scattering in liquid neon.
\subsection[Dark matter rates]{Dark matter recoil spectra and rates}
The differential nuclear recoil spectrum for WIMP dark matter elastic scattering from  a target with nuclear mass A is given by  (in events per unit energy per unit time per unit detector mass)~\cite{jungman}
\begin{equation}
\frac{dR}{dQ}  =  \frac{\sigma_{0} \rho_{0}}{\sqrt(\pi)v_{0}m_{\chi}m_{r}^2} F^{2}(Q)T(Q) \label{eqn:wimp_recoil}
\end{equation}

\noindent with
\begin{eqnarray*}
Q & = & \mbox{unquenched nuclear recoil} \\
  &   & \mbox{kinetic energy} \\
\sigma_{0} & = & \mbox{WIMP-nucleus ES cross section} \\
\rho_{0} & = & \mbox{local dark matter density} \\
v_{0} & = & \mbox{speed of Sun around galactic center} \\
m_{\chi} & = & \mbox{WIMP mass} \\
m_{r} & = & \mbox{reduced WIMP-nucleus mass} \\
      & = &  \frac{m_{\chi}m_{N}}{m_{\chi}+m_{N}} \\
m_{N} & = & \mbox{Nuclear mass} \\
F(Q) &=& \mbox{form factor} \\
T(Q) & = & \frac{\sqrt(\pi)}{4v_{e}}v_{0} \times \\
  &&\left[\rm{erf}\frac{v_{min}+v_{e}}{v_{0}}-\rm{erf}\frac{v_{min}-v_{e}}{v_{0}}\right] \\
v_{min} & = & \mbox{minimum WIMP velocity} \\
v_{e} & = & \mbox{earth's velocity.}
\end{eqnarray*}

We use here the standard nuclear form factor
\begin{equation}
F(Q) = \left[ \frac{3j_{1}(qr_n)}{qr_n}\exp[-\frac{1}{2}(qs)^2] \right]
\label{form_factor}
\end{equation}
\noindent with the numerical values given by ~\cite{lewin}:

\begin{eqnarray*}
qr_n & = & 6.92\times10^{-3}A^{1/2}Q^{1/2}(1.14A^{1/3}) \\
qs   & = & 6.92\times10^{-3}A^{1/2}Q^{1/2}{0.9} \mbox{, }
\end{eqnarray*}

\noindent $j1$ is spherical Bessel function, and $Q$ is the nuclear recoil energy in keV.

To compare experiments with different target nuclei, we express the nuclear cross-section in terms of the nucleon cross-section~\cite{bottino}

\begin{equation}
\sigma_{0} = \sigma_{P}\left[ \frac{1+m_{\chi}/m_{N}}{1+m_{\chi}/m_{p}} \right]^2 A^{2}
\label{eqn:sigma0}
\end{equation}
\noindent with
\begin{equation}
m_{p}  =  \mbox{proton mass}
\end{equation}
\noindent where we consider only the spin-independent interaction.

\begin{figure}[h]
\begin{center}
\includegraphics[width=\figwidth]{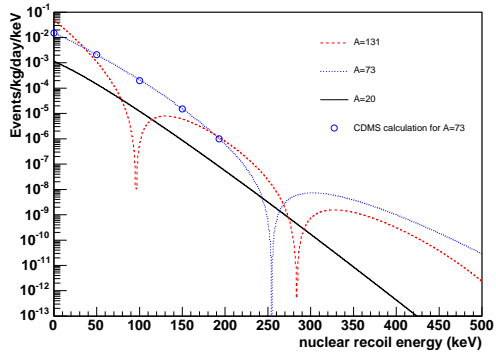}
\caption[Nuclear recoil spectra from WIMPs]{\label{wimp_rates}Nuclear recoil spectra (unquenched kinetic energy) from WIMP elastic scattering on Xe (A=131), Ge (A=73) and Ne (A=20) calculated with equation~\ref{eqn:wimp_recoil}.  Rates have been calculated assuming $\sigma_{P}$ = $10^{-42}$ cm$^{-2}$ and m$_{\chi}$ = 100 GeV, and are in units of keV$^{-1}$kg$^{-1}$day$^{-1}$.  The points are from a calculation by the CDMS collaboration~\cite{cdms_longpaper}.}
\end{center}
\end{figure}

\begin{figure}[h]
\begin{center}
\includegraphics[width=\figwidth]{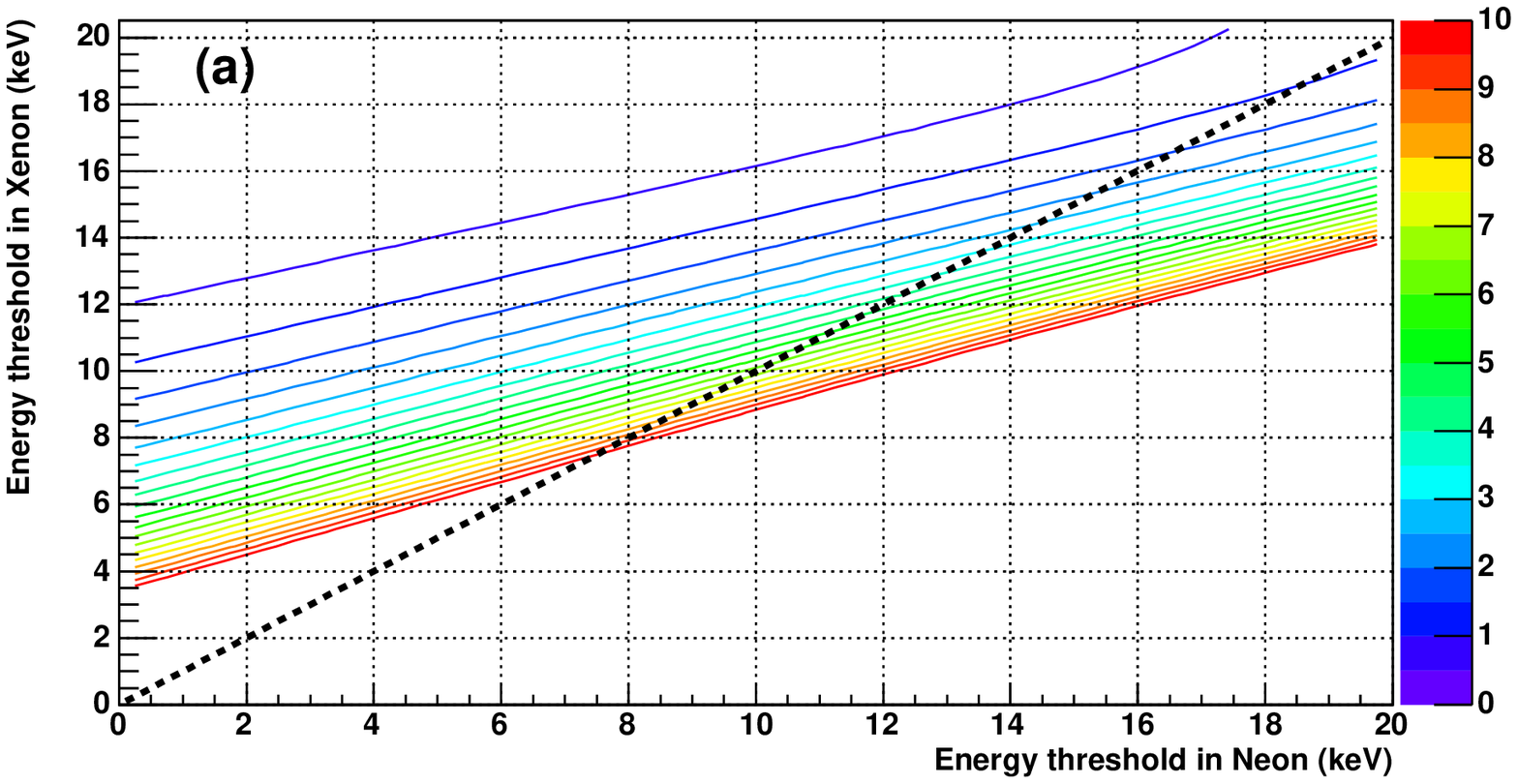}
\includegraphics[width=\figwidth]{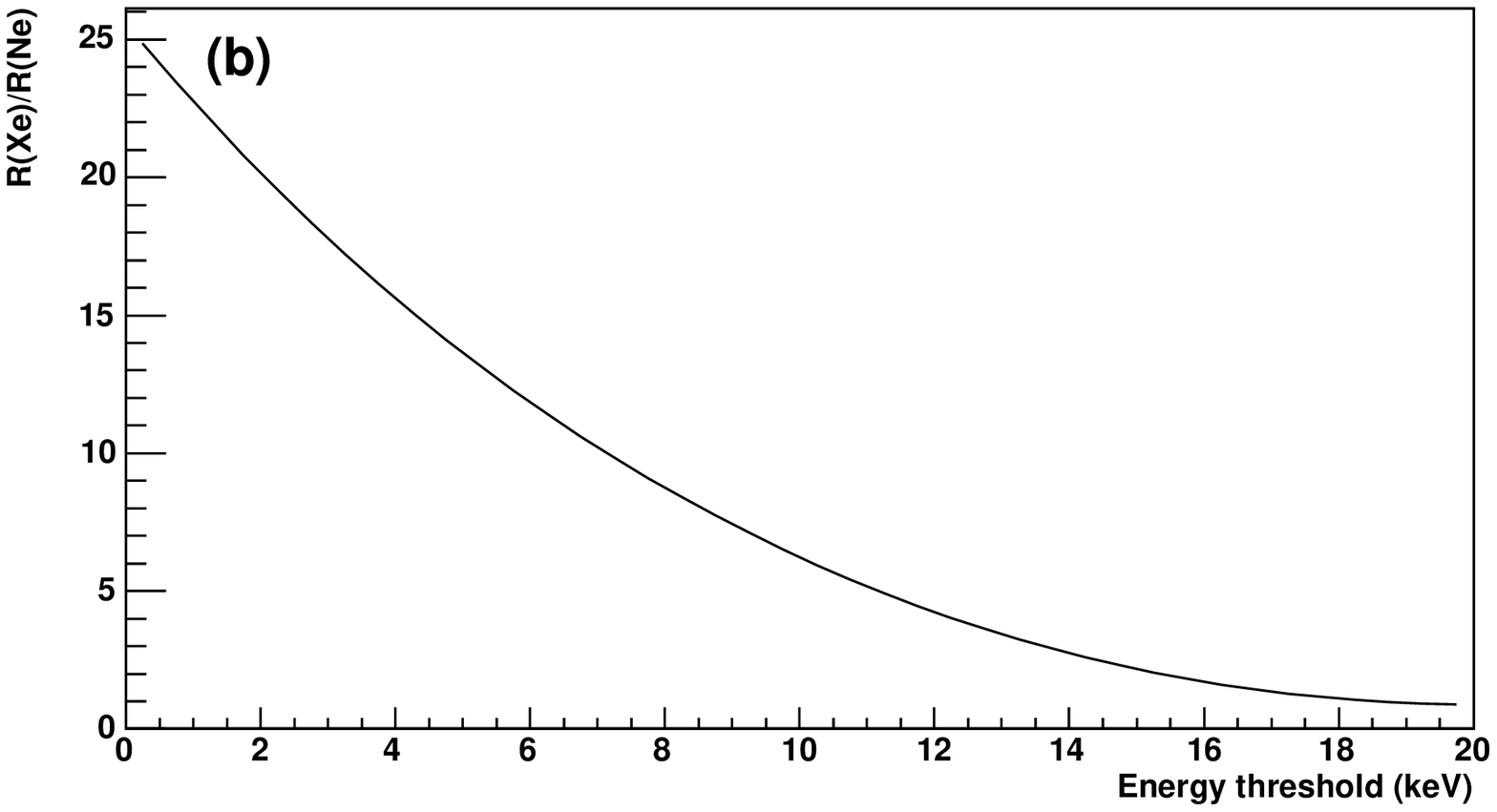}
\caption{\label{compare_xe_ne}(a) Relative rates above threshold per unit mass of material for a 100 GeV WIMP on Xe and Ne.  Contours show relative rates above
the effective energy threshold, assuming 25\% quenching for Xe and Ne.  (b)  Relative rate above threshold assuming the same quenching (25\%) and threshold for both Xe and Ne.
The relative rate decreases rapidly with increasing threshold, and is approximately 6 for a 10 keV threshold.} 
\end{center}
\end{figure}

Figure~\ref{wimp_rates} shows the nuclear recoil spectrum calculated with equation~\ref{eqn:wimp_recoil}
for the input parameters defined in Table~\ref{wimp_pars_table} for several target nuclei.   A potential advantage over
a xenon-based detector is the lower sensitivity to the achievable energy threshold due to the shape of the recoil spectrum.  Figure~\ref{compare_xe_ne}
shows the ratio of expected dark matter rates above threshold for xenon and neon targets, where it is seen that
the relative xenon sensitivity drops quickly as the energy threshold is increased.  The nuclear form factor correction (Eqn.~\ref{form_factor}) tends to suppress the predicted event rate for large A nuclei with increasing kinetic energy, so that for experimentally achievable energy thresholds, much of the gain obtained by exploiting the coherence predicted by equation~\ref{eqn:sigma0} for large-A target nuclei is lost.
 With a 10 keV threshold, for example, the additional sensitivity with xenon is only approximately a factor of 6, which can be overcome with neon by
using a larger target mass.

\begin{table}[htb]
\begin{center}
\caption[Parameters used for WIMP recoil calculation]{\label{wimp_pars_table}Parameters used for calculation WIMP recoil spectrum and rates.}
\begin{tabular}{ll} \hline
Parameter & Value \\ \hline
$\rho_{0}$ & 0.3 GeV c$^{-2}$ \\
$v_{0}$ & 220 km s$^{-1}$ \\ \hline 
\end{tabular}
\end{center}
\end{table}

\subsection[Nuclear and e$^{-}$ separation]{Distinguishing nuclear and electron recoils in neon}
\begin{figure}[h]
\begin{center}
\includegraphics[width=\figwidth]{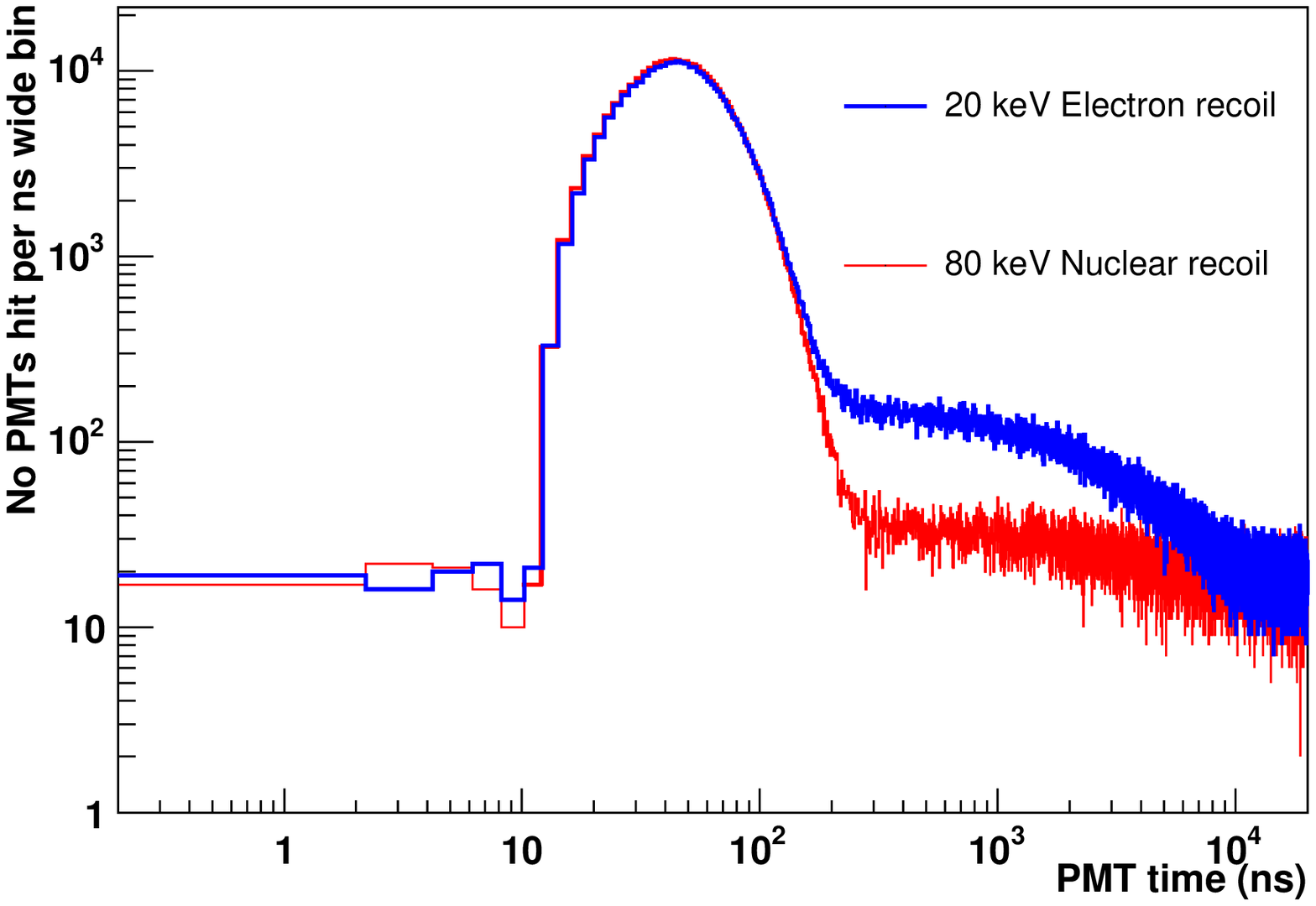}
\includegraphics[width=\figwidth]{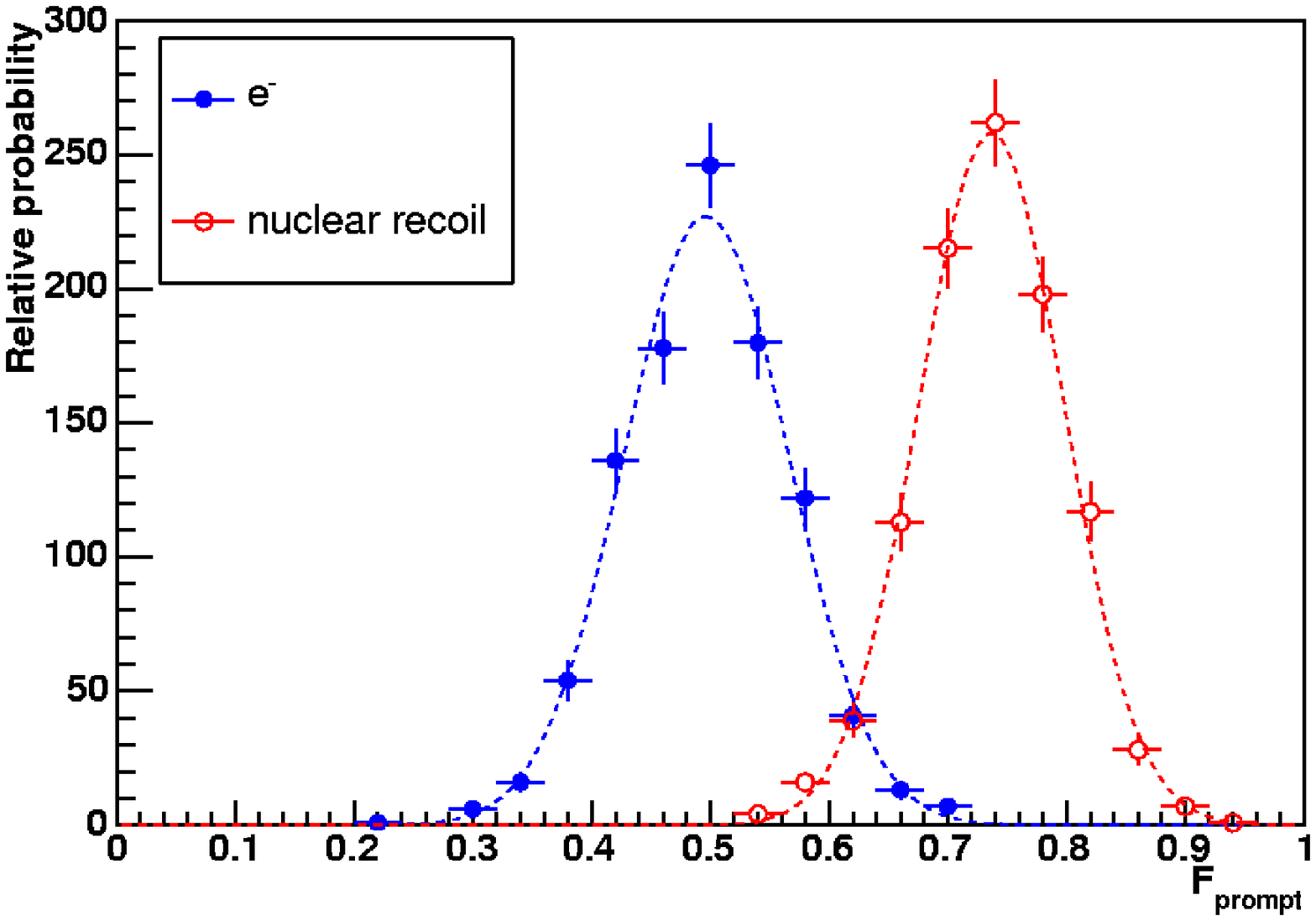}
\caption[Distinguishing electron from nuclear recoils]{\label{wimp_vs_e}Distinguishing electron versus nuclear recoil events based on PMT time distribution.  (top) PMT time distribution for 10000 simulated electron recoil and nuclear recoil events. (bottom) Distribution of $f_{prompt}$ for these sets of events.}
\end{center}
\end{figure}

Assuming we have designed the experiment so that the backgrounds to the $pp$ neutrinos after threshold and fiducial volume cuts are small,
the dominant background to potential dark matter events will arise from the $pp$ neutrinos themselves.  In this case the ultimate sensitivity to dark matter will 
be limited by the expected dark matter interaction rate and the separability of dark matter events from neutrino events.  The PMT time distribution can in principle
be used to separate these classes of events, since the intrinsic scintillation time distribution is different for electron and nuclear recoils (see Table~\ref{table_quench}).  The PMT time distribution predicted by the Monte-Carlo for 20 keV electron events and 80 keV nuclear recoil events from WIMP elastic scattering is shown in Fig.~\ref{wimp_vs_e}(a).  These events
would yield similar effective energy distributions due to the quenching of the scintillation light from nuclear recoils.  To separate these classes of events,
the ratio $f_{prompt}$ of prompt to total hits is calculated.  These distributions are shown in Fig.~\ref{wimp_vs_e}(b).  The effect of PMT noise is to decrease
the prompt fraction.

We can now use these distributions to statistically separate nuclear recoils from electron events.  The separation efficiency was estimated by generating 1000 MC datasets, each with 3000 solar neutrino events, corresponding to approximately one live year.
We define the extended negative likelihood function
\begin{equation}\begin{split}
\mathcal{L} &= -2(N_{e}+N_{\chi}) \sum_{i=1}^{N_{events}} [ N_{e}P_{e}(f_{prompt}^{i}) + \\
 &N_{\chi}P_{\chi}(f_{prompt}^{i}) ]  \label{max_like}
\end{split}\end{equation}
with $P_{e}$ and $P_{\chi}$ the probability distributions for $f_{prompt}$ shown in Fig.~\ref{wimp_vs_e}(b).  We can then minimize equation~\ref{max_like} to find
the most likely values of $N_{e}$ and $N_{\chi}$.  Since equation~\ref{max_like} is an extended likelihood, the $s-\sigma$ parameter uncertainties can be calculated by finding the iso-likelihood contours which satisfy
\begin{equation}
\mathcal{L}-\mathcal{L}_{min} = s^{2}
\end{equation}
\noindent and include the sampling uncertainties in our dataset, so that they include the uncertainties on deduced incident particle fluxes.

\begin{figure}
\begin{center}
\includegraphics[width=\figwidth]{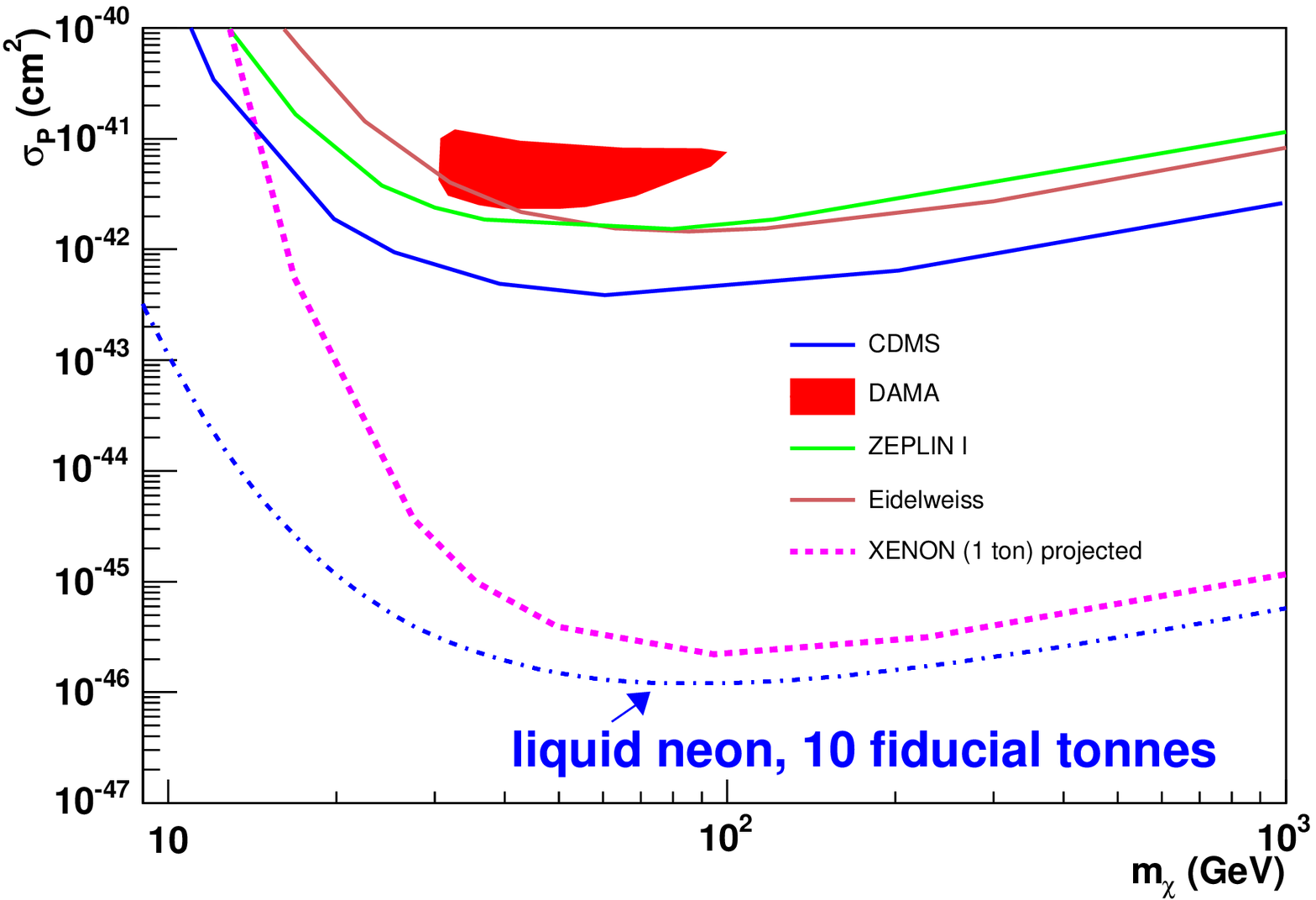}
\caption[Projected limits on WIMP-nucleon cross-section]{\label{wimp_limits}Limits on the WIMP-nucleon spin-independent cross-section versus WIMP mass for current and planned experiments and projected sensitivity for liquid neon.  Data from current experiments and projected XENON sensitivity from Gaitskell and Mandic~\cite{dmtools}.}
\end{center}
\end{figure}

Equation~\ref{max_like} was minimized (with the ROOT TMinuit package~\cite{root}) for each of the 1000 simulated datasets.  The fitted numbers of electron and WIMP events were found to peak at the input values, providing a check of signal analysis.  The 90\% CL upper limit on the number of WIMP events in the data set is estimated from this technique to be approximately 1.3 events, which provides background discrimination of approximately 0.9965.  We can use this upper limit along with the predicted nuclear recoil rate from WIMP scattering (which is the convolution of
equation~\ref{eqn:wimp_recoil} with the energy response function of Fig.~\ref{teff_res}, corrected for 25\% quenching) to plot the sensitivity contour in WIMP-nucleus cross-section versus WIMP mass.  This is shown in Fig.~\ref{wimp_limits}, along with current and projected experimental results.  The limit corresponds to a WIMP-nucleon spin-independent cross section ($\sigma_{p}$) of $\sim$$10^{-46}$ cm$^{2}$ for a 100 GeV WIMP mass.  Details on the selection of optimal analysis windows and cuts are provided in the following section.

\subsection[Optimizing sensitivity]{Optimizing detector sensitivity}

\begin{figure}[h]
\begin{center}
\includegraphics[width=\figwidth]{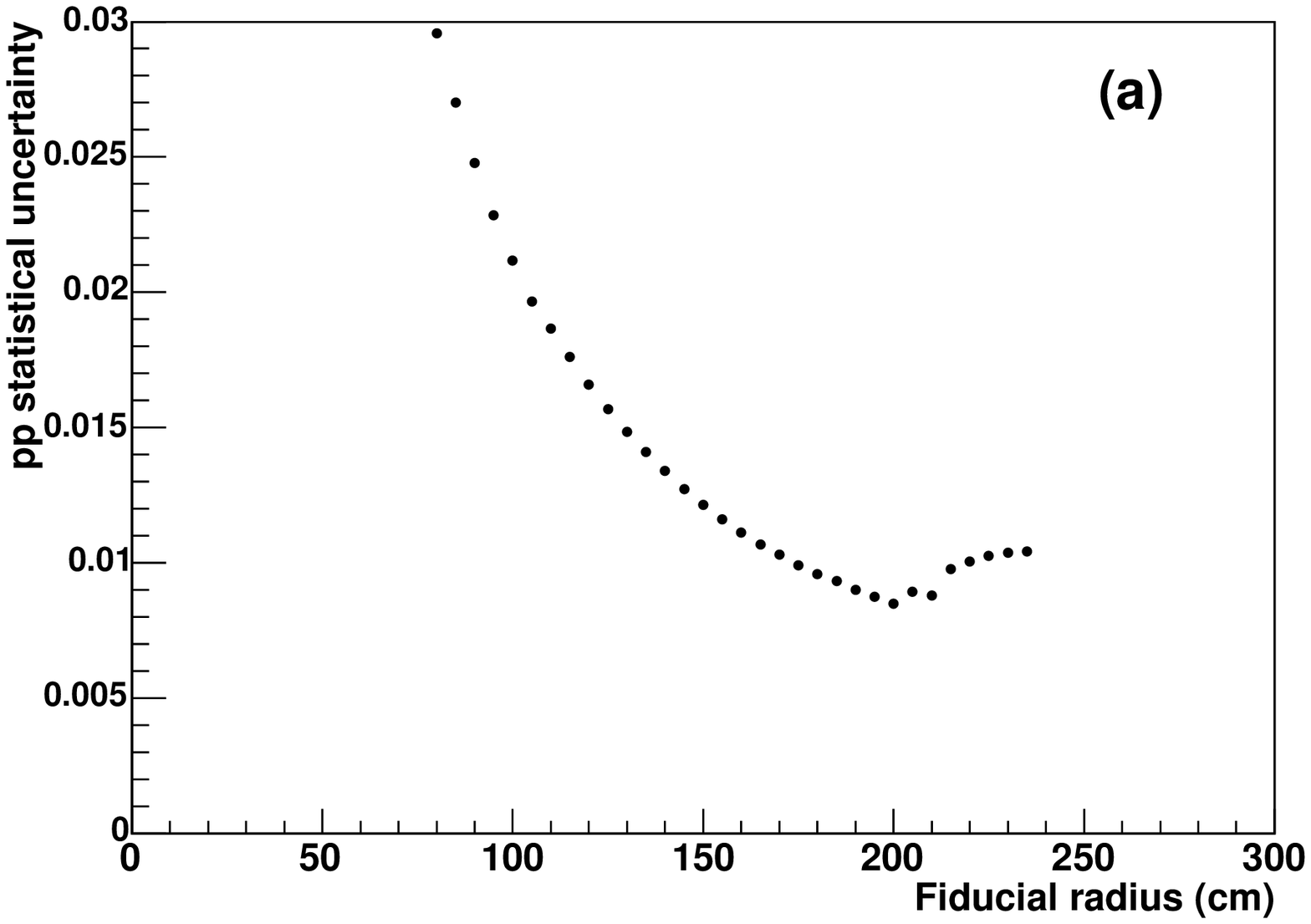}
\includegraphics[width=\figwidth]{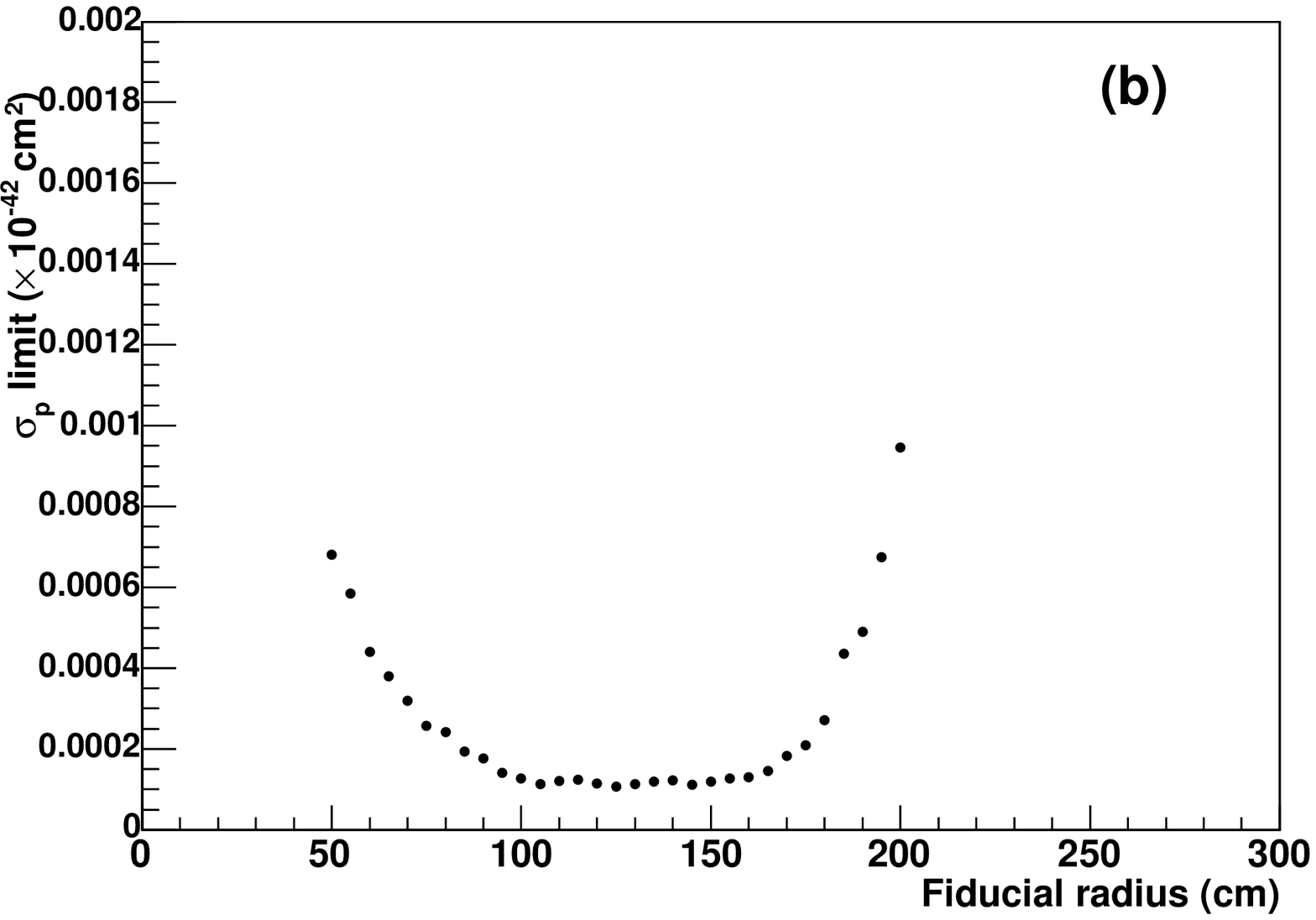}
\caption{\label{optimal_300}Optimal fiducial volume determination for the $pp$ neutrino (a) and dark matter (b) experiments for a nominal PMT sphere radius of 300 cm, and 1 year of livetime.}
\end{center}
\end{figure}

\begin{figure}[h]
\begin{center}
\includegraphics[width=\figwidth]{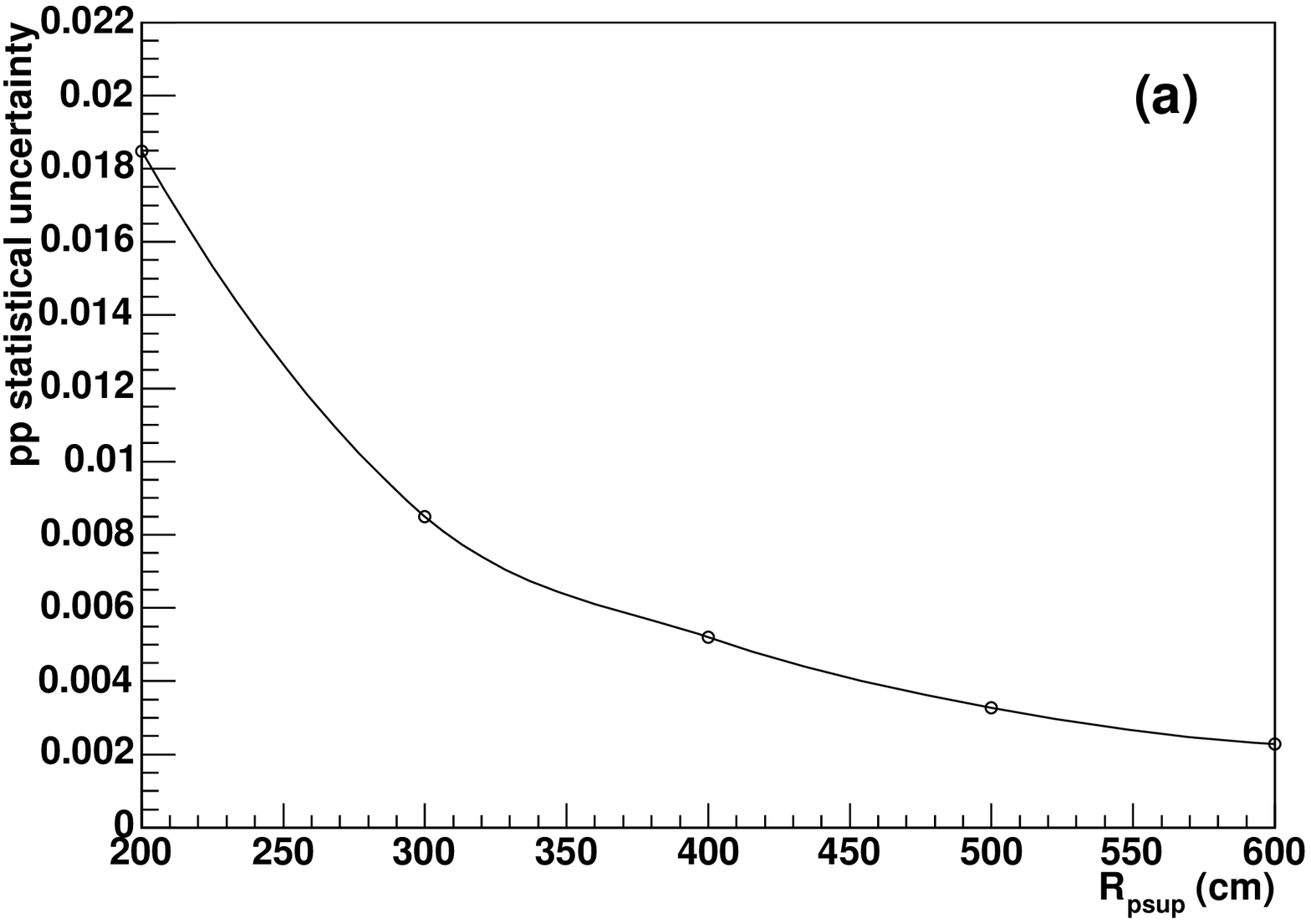}
\includegraphics[width=\figwidth]{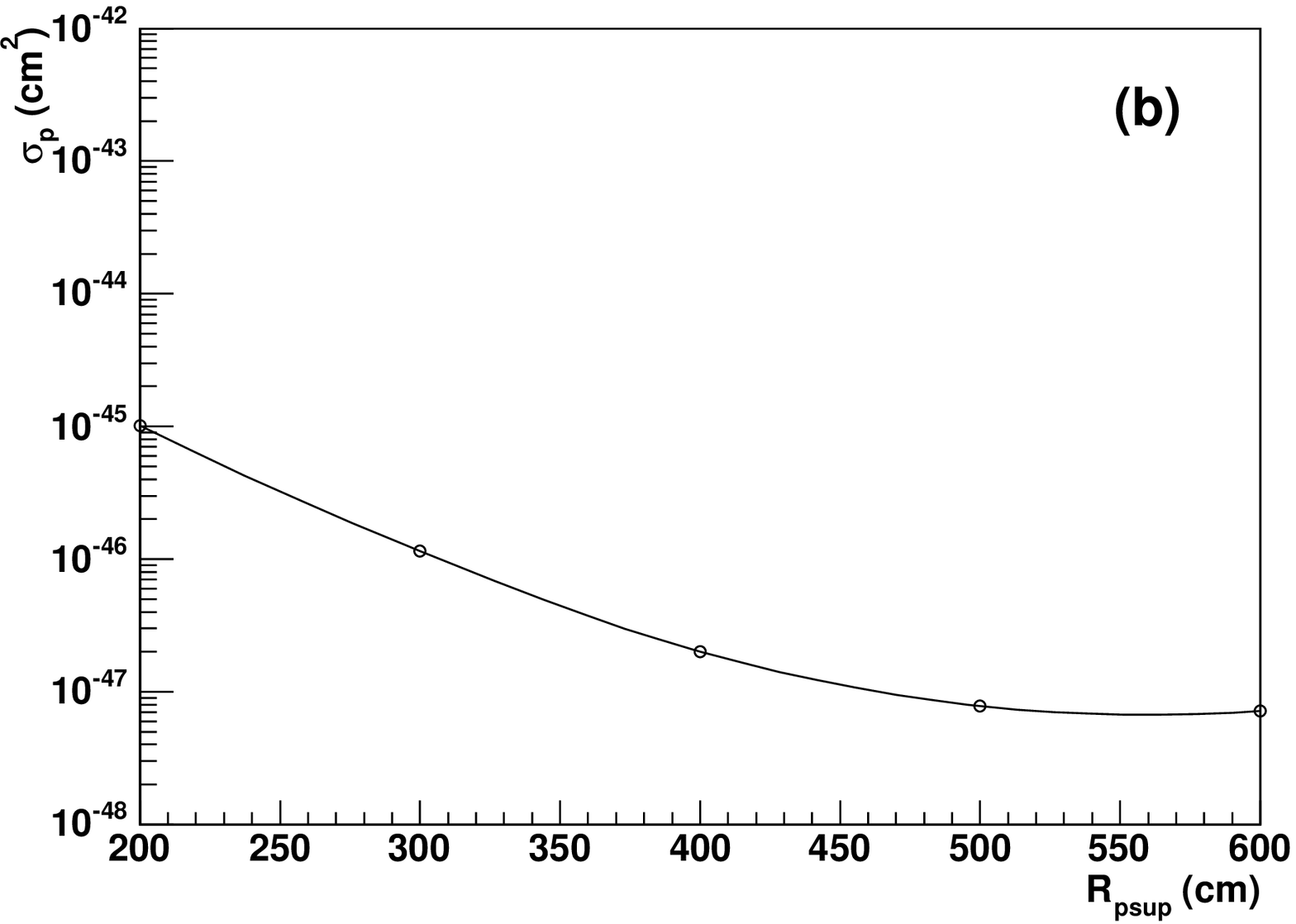}
\caption{\label{sensvsrpsup}Sensitivity for (a) the $pp$ neutrino and (b) dark matter experiments versus radius of PMT sphere (R$_{\rm psup}$).  The fiducial volume and thresholds are optimized for each value of R$_{\rm psup}$.}
\end{center}
\end{figure}

The optimal energy threshold and fiducial volume for the dark matter and $pp$ experiment will in general be different, since the energy spectra
of these two event classes differ.  To simultaneously optimize the experiment for dark matter and $pp$ neutrinos, an optimum fiducial volume
and energy threshold are found for each by maximizing the total predicted event rate above threshold as a function of fiducial volume cut.
It was found that an energy window of 15 keV above threshold provided the best WIMP sensitivity for a 100 GeV WIMP, and this window was employed
throughout the analysis.
Fig.~\ref{optimal_300} shows the determination of the optimal fiducial volume for neutrinos and dark matter for a 300 cm detector.  It is found that the sensitivity for dark matter is optimized
by selecting a smaller fiducial volume and thus a  lower analysis threshold than for the solar neutrino analysis, since the energy spectrum of the WIMP
interactions is much harder.  The optimized thresholds and fiducial volumes, which are different for the dark matter and $pp$ components of the experiment, have
been used in the analysis.  From Fig.~\ref{optimal_300}(b), the relatively large region over which the dark matter sensitivity is optimized is due to a trade-off
between a lower analysis threshold (for smaller fiducial radii) and better WIMP-neutrino separation with increasing analysis threshold (for larger fiducial radii). 
This effect can be used a systematics check on the data from the liquid neon experiment and allows some flexibility in ultimately determining the analysis cuts.
The optimized sensitivities for the $pp$ and dark matter experiments are shown versus detector size in Fig.~\ref{sensvsrpsup}.  From this figure a 300 cm radius detector
is found to provide good sensitivity for both measurements.  Little additional sensitivity is found for detector sizes much larger than 400 cm in radii, since
the measurements become limited by position resolution.

\clearpage 

\section{Detector calibration requirements}
\label{section_calibration}
Assuming radioactive backgrounds to the $pp$ flux measurement can be reduced such that they contribute less than $1\%$ uncertainty,
the dominant systematic uncertainties will be due to uncertainty of the acceptance of the solar neutrino signal.  These uncertainties
will arise through uncertainty of the energy threshold and fiducial target mass.
Table~\ref{pp_error_budget} outlines a nominal error budget assuming a target $pp$ flux uncertainty of 1\%, where the dominant systematic uncertainties are assumed to 
each contribute an equal 0.35\% to the total uncertainty.
The dominant fiducial mass (volume) uncertainty arises from uncertainty in the difference between the true and reconstructed radii of events.  The fractional radial uncertainty $\epsilon_{R}=\frac{\Delta R}{R}$ is related to the fiducial target uncertainty $\frac{\Delta V}{V}$ by 
\begin{equation}
\epsilon_{R} = \sqrt[3]{\frac{\Delta V}{V}+1}-1 .
\end{equation}
The requirement of a 2.5 mm position uncertainty for a 200 cm fiducial volume will require a calibration source which can be positioned to that accuracy, an order
of magnitude improvement over that currently achieved by SNO and similar experiments.  This requirement could be relaxed if the overall uncertainty on the $pp$ flux is allowed to increase, if other systematic uncertainties are lower than shown in Table~\ref{syst_uncert}, or if the statistical uncertainty is reduced by increasing detector livetime. 

\begin{figure}[h]
\begin{center}
\includegraphics[width=\figwidth]{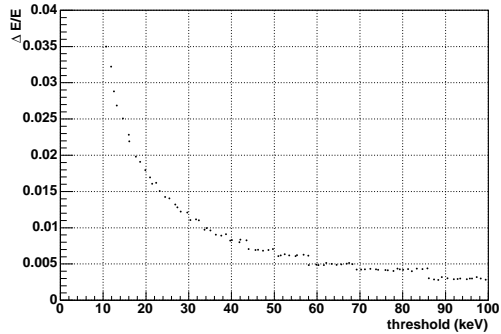}
\caption[Energy calibration requirements]{\label{thresh_uncert}Absolute energy scale calibration uncertainty (as a fraction of the threshold energy) versus
threshold for a $0.35\%$ systematic uncertainty on the measured $pp$ flux.}
\end{center}
\end{figure}

 The uncertainty due to the absolute energy scale calibration can be evaluated by calculating the change in area of the spectrum of Fig.~\ref{nu_spec}(b) when the threshold
changes by one standard error.  The minimum threshold uncertainty versus threshold is shown in Fig.~\ref{thresh_uncert} for the target  $pp$ flux systematic uncertainty due to energy scale calibration of $0.35\%$.
  Shown in Table~\ref{syst_uncert} are the requirements
for calibration of the absolute energy scale (near threshold) and the position reconstruction offset for a resultant $0.35\%$ uncertainty on the $pp$ flux, assuming a 35 keV threshold.  It is interesting to note that a large scintillation yield versus neon temperature dependence [approx ($8\%$/K)] is found in~\cite{neon_scint}, which leads to the requirement that the neon be temperature stable at the level of $\Delta$T$<125$~mK to satistify the requirements of Table~\ref{syst_uncert}.

The calibration requirements for the dark matter component of the experiment are much less stringent.  Fiducial volume uncertainties of a few percent achieved
by current generation experiments are adequate.  An uncertainty of 20-30\% in the energy scale calibration for nuclear recoil events near the threshold ($\approx$ 10 keV)
leads to a factor of two change in the deduced WIMP cross-section, and will be achievable with calibration by a neutron source.

\begin{table}[htb]
\begin{center}
\caption{\label{pp_error_budget}Error budget for pp flux measurement with R$_{\rm psup}$=300 cm, 1 year livetime.}
\begin{tabular}{ll} \hline
Source             & Relative uncert.  \\ 
                   & $(\%)$                \\ \hline
statistics         & 0.7                  \\
energy scale       & 0.35                   \\
vertex accuracy    & 0.35                    \\
uranium background & 0.35                    \\
thorium background & 0.35                    \\  
krypton background & 0.35                    \\ \hline
{\bf{total}}              & 1.0                   \\ \hline
\end{tabular}
\end{center}
\end{table}

\begin{table}[htb]
\begin{center}
\caption[Energy and position calibration requirements]{\label{syst_uncert}Absolute energy scale and position calibration requirements for $0.35\%$ contributions
to the $pp$ flux systematic uncertainty, assuming a threshold of 35 keV. }
\begin{tabular}{l|l} \hline
energy scale ($\Delta$E/E) & $<1 \%$ \\ 
vertex accuracy ($\Delta$R/R)& $<0.12\%$ \\
vertex accuracy (at R=200 cm) & 2.5 mm \\ \hline
\end{tabular}
\end{center}
\end{table}

\section{Design criteria}
Table~\ref{response_req} summarizes measurements which must be made to allow for an optimized design of a liquid neon $pp$ neutrino and dark matter experiment.
An ``X'' in either the dark matter or $pp$ neutrino column denotes the measurement primarily affects this component of the experiment.
Dependence of the ultimate detector sensitivity on these parameters is discussed in the following sections.

\begin{table*}[htb]
\begin{center}
\caption[Response requirements for design criteria]{\label{response_req}Requirements for design criteria of a liquid neon detector for both solar neutrinos and dark matter.  Response characteristics
marked with ``X'' need to be measured (or shown to be within range) before the detector design can be optimized for either the dark
matter or solar neutrino experiment.}
\begin{tabular}{llll} \hline
Response & Dark matter & Solar $\nu$ \\ \hline
e$^-$ scint. yield  & X & X \\ 
e$^-$ scint. prompt/late ratio&X & \\
Scintillation yield linearity & X & X \\
nucleus scint.  yield & X &  \\
nucleus scint. prompt/late ratio & X & \\ 
late WS time constant &X & \\
Rayleigh scattering & & \\
absorption length &X &X \\
neon temperature stability &  &X \\
PMT long-term stability & X & X \\
PMT QE &X &X \\
PMT threshold & X & X \\ 
PMT dark noise &X &X \\ \hline
\end{tabular}
\end{center}
\end{table*}

\subsection[Photon absorption in neon]{Photon absorption in liquid neon}

\begin{figure}[h]
\begin{center}
\includegraphics[width=\figwidth]{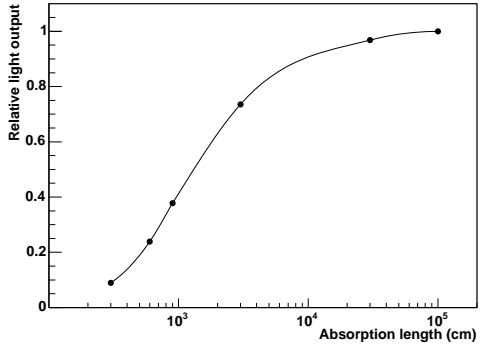}
\caption{\label{light_output}Relative light output (normalized to no absorption) versus absorption length for events at the detector center.}
\end{center}
\end{figure}

Absorption of the primary scintillation photons in the liquid neon will both
reduce the total light output and introduce systematic effects related to 
event position and direction.  To estimate the effects of absorption on total light
output, a set of Monte-Carlo calculations were performed for a range of photon
absorption lengths.  Fig.~\ref{light_output} shows the light output (relative to no absorption) for scintillation events at the center of a 300 cm radius detector for absorption lengths ranging from 100 to 10$^{5}$ cm.  The effect of the 60 cm Rayleigh scattering length is to greatly increase the optical path length for photons, and thus decrease the tolerable  amount of absorption.  To obtain a relative light yield of 95\%, an absorption length of approximately 
300 m is required.

The absorption coefficient ($\alpha$) for scintillation photons can be estimated by adding the absorption coefficients for the various absorbers present in the liquid:

\begin{equation}
\alpha = \frac{1}{\lambda} = \sum_{i}^{N} \frac{\rho N_{A} \gamma_{i} \sigma_{i}}{M_{i}}
\end{equation}
\noindent where
\begin{eqnarray*}
M_{i} & = & \mbox{Molar mass of absorber} \\
\rho & = & \mbox{liquid neon density} \\
N_{A} & = & \mbox{Avogadro's constant} \\
\gamma_{i} & = & \mbox{Mass fraction of absorber} \\
\sigma_{i} & = & \mbox{absorption cross section}
\end{eqnarray*} 

If we assume the absorption is equally divided among N absorbing materials\footnote{This is likely an oversimplification since some materials will pose more of a problem than others, but this nonetheless provides an estimate of the tolerable contaminations.}, we can set the mass fraction limit ($\gamma_{i}$) for 
each absorber:
\begin{equation}
\gamma_{i}  =  \frac{M_{i}}{N \rho N_{A} \sigma_{i} \lambda}
\end{equation}

Neon has no significant absorption for photons with wavelengths above the first ionization limit at 575 nm, and so is transparent to its scintillation
light \cite{lee}. A comprehensive list of photo absorption cross-sections has been compiled by Gallagher {\em{et. al.}} \cite{gallagher}.  Shown in Table~\ref{absorption_table} is a list of dominant absorbers present in neon, along with the cross-sections for photo absorption of 80 nm photons from~\cite{gallagher},
and the required impurity limits for a total photon absorption length of 300 m.  These levels are moderate compared to that  required for ${}^{85}$Kr, and most impurities are expected to be easily removed from the neon through the use of cold traps.  The problem of hydrogen diffusion into the neon is expected to be mitigated by the very low diffusion rate at liquid neon temperature.

\begin{table}[h]
\begin{center}
\caption{\label{absorption_table}Total absorption cross sections for 80 nm photons and impurity limits for a 300 m absorption length.}
\begin{tabular}{lll} \hline
Absorber & Cross section at 80 nm & $\gamma_{i}$ \\ 
         & ($\times 10^{-18}$ cm$^{2}$) & ($\times 10^{-12}$)  \\ \hline
H$_{2}$    & 10.5  &   2.2      \\
H$_{2}$O   & 21    &   9.8      \\ 
N$_{2}$    & 12    &   26.8     \\ 
O$_{2}$    & 17    &   21.6     \\ \hline
\end{tabular}
\end{center}
\end{table}

\subsection[Dependence on inputs]{Dependence on optical and detector parameters}
\label{sensitivity_section}
\begin{figure*}[h]
\begin{center}
\includegraphics[width=\figwidthtwo]{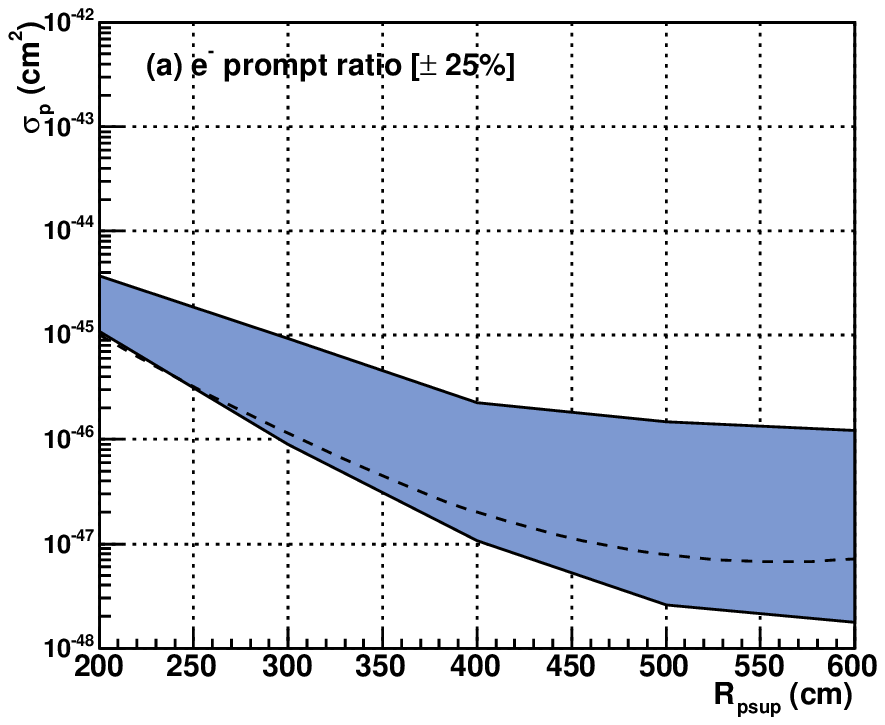}
\includegraphics[width=\figwidthtwo]{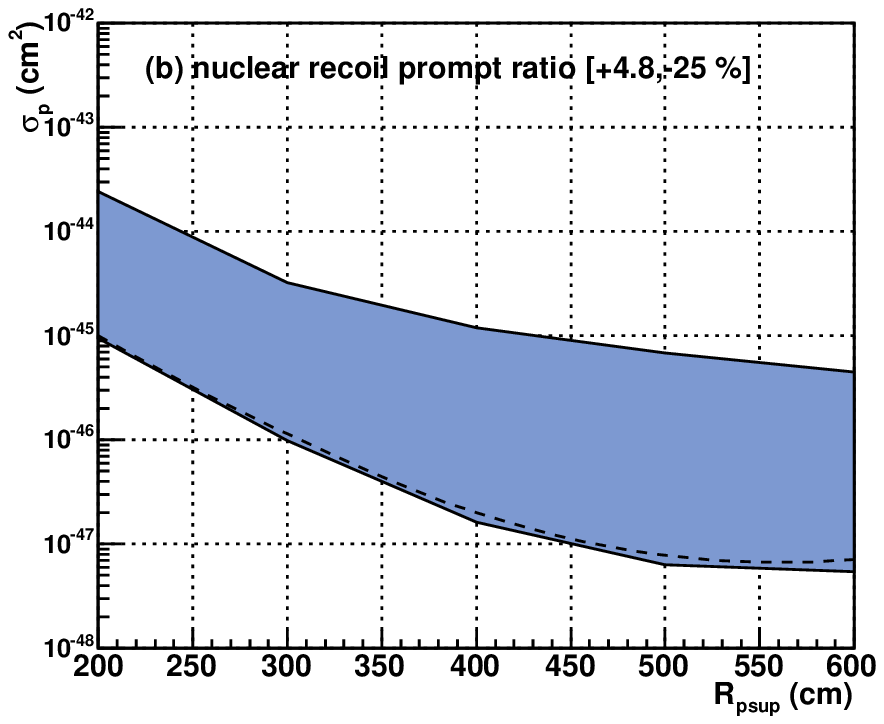}
\includegraphics[width=\figwidthtwo]{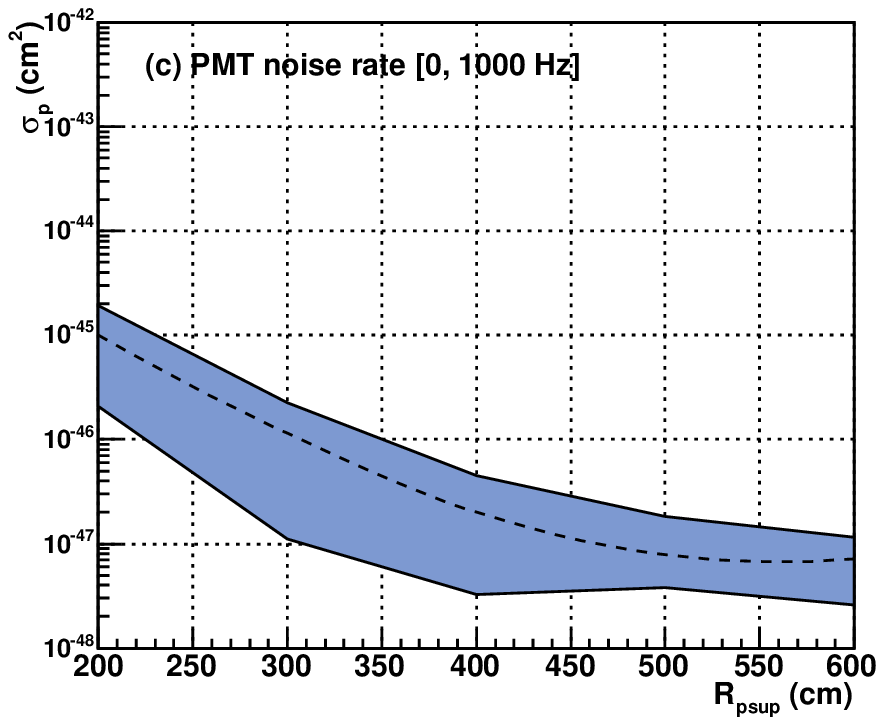}
\includegraphics[width=\figwidthtwo]{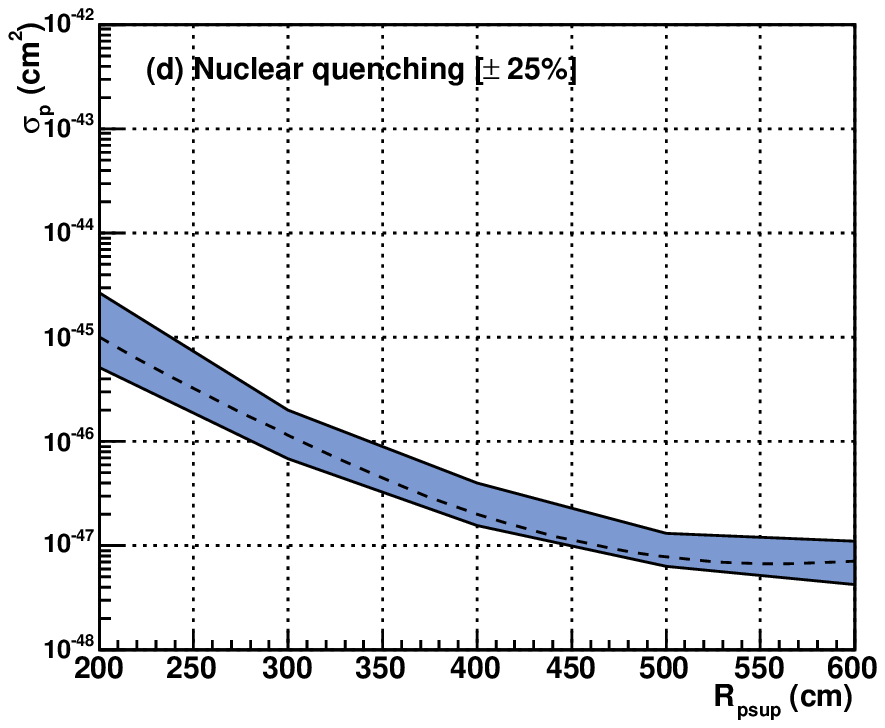}
\includegraphics[width=\figwidthtwo]{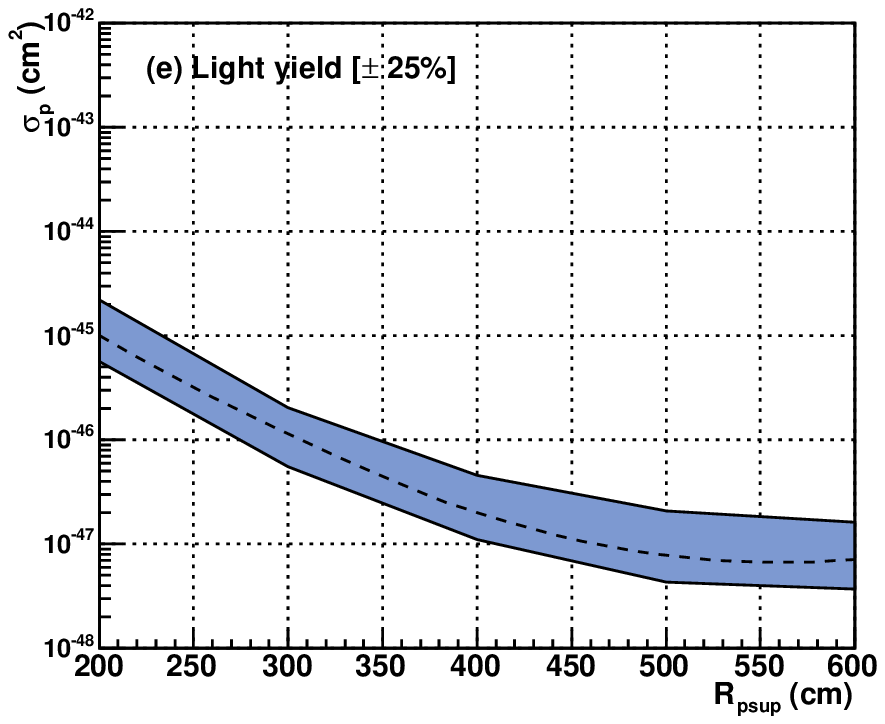}
\includegraphics[width=\figwidthtwo]{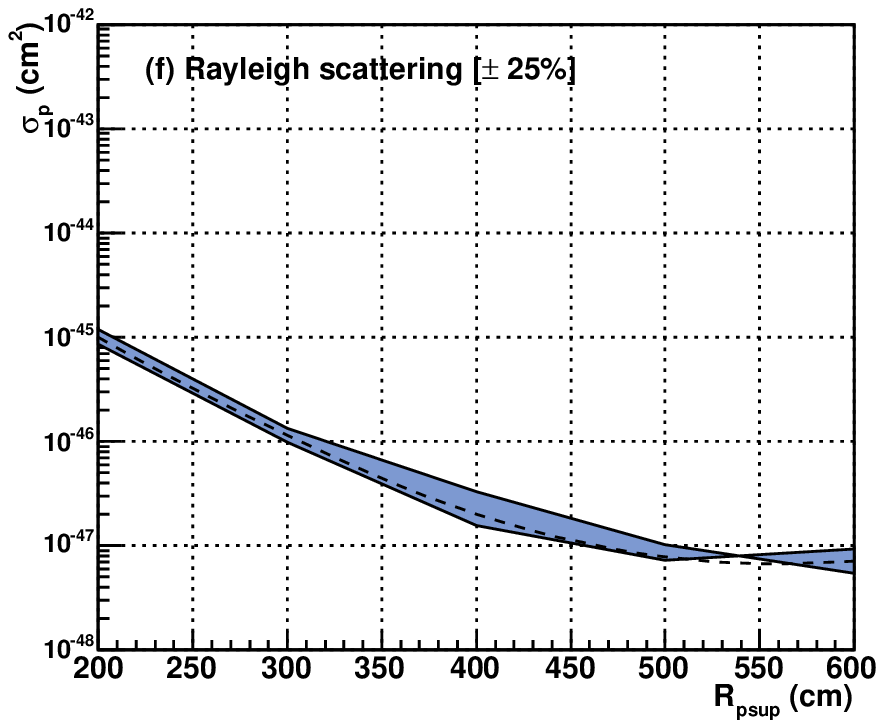}
\caption{\label{par_dep}Dependence of WIMP sensitivity on optical and detector parameters.  Central values are from model simulations and are shown as dashed curves.  Shown are the ranges in sensitivity for (a) electron prompt to total light ratio varied by $\pm 25\%$, (b) nuclear recoil prompt to total light ratio varied by +4.8,-25$\%$, (c) PMT noise rates of 0 and 1000 Hz, (d) Light quenching factor for
nuclear recoils varied by $\pm 25\%$, (e) total scintillation yield varied by $\pm 25\%$, and (f) Rayleigh scattering varied by $\pm 25\%$.  The WIMP sensitivity depends most critically on the prompt to total light ratio for electrons and
nuclear recoils.}
\end{center}
\end{figure*}

Many of the parameters shown in Tables~\ref{neon_scint_prop}~and~\ref{table_quench} are either estimates inferred from other measurements or have not been quantified with systematic uncertainties.  The sensitivity of the solar $pp$ neutrino measurement to these parameters is quite modest.
To study the sensitivity of the dark matter experiment to these inputs, a set of Monte-Carlo calculations were performed by varying the parameters
which will affect the detector performance by a nominal amount ($\pm 25\%$ in most cases, or a physical bound).  Figure~\ref{par_dep} shows the deduced
sensitivity for a 100 GeV WIMP versus detector size, for the ranges of optical and detector parameters of interest.  This provides a benchmark 
for the required accuracy with which these parameters need to be measured, and points out the parameters to which we are most sensitive.  The prompt to total
light yield for electrons and nuclear recoil events provide separation between these two classes of events, and are found to be the dominant 
factors in determining the ultimate WIMP sensitivity.   It is interesting to note that the  WIMP sensitivity varies only weakly with
total light yield, nuclear recoil quenching factor, and the Rayleigh scattering.  This is somewhat fortuitous since measuring these absolute quantities
experimentally in the laboratory with high precision is difficult.  Experiments are currently underway at Los Alamos National Laboratory and Yale University to measure these properties.

\begin{figure}[h]
\begin{center}
\includegraphics[width=\figwidth]{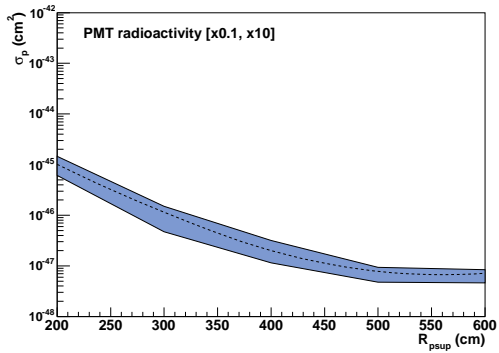}
\caption{\label{pmt_back_shift}WIMP sensitivity for an order of magnitude change in the PMT backgrounds.}
\end{center}
\end{figure}

Figure~\ref{pmt_back_shift} shows the effect of changing the total PMT radioactivity by $\pm$ one order of magnitude from the nominal values
for PMTs with commercially available ultra-low background glass.  This is found to also have a relatively small effect on the estimated WIMP sensitivity.

\clearpage

\section{Discussion}

In summary, we have demonstrated the possibility for a simultaneous dark matter and $pp$ neutrino experiment using
liquid neon, assuming nominal scintillation characteristics and background contamination levels with detailed
Monte-Carlo calculations.  This approach to direct dark matter detection has the advantage that
WIMP event discrimination is based only on PMT signals, allowing for a simple detector design which can 
in principle be scaled to a very large target mass.
 This detector design also has the potential advantage that external sources of background are now well-separated
from the inner target material, and can be dealt with through position reconstruction.  Assuming light attenuation
and internal source radioactive backgrounds can be controlled, the large target mass achievable
with neon may lead to the best sensitivity of any of the direct dark matter searches.
We have shown that form factor suppression for high-A target nuclei leads to less of an advantage over lower-A target
nuclei than that expected by the cross-section coherence alone.  A specific list of issues which must be addressed in order
to determine the ultimate detector sensitivity has been provided.  An experimental program is underway to address these issues.

\section*{Acknowledgments}

This work was supported with funds from the Los Alamos Directed Research and Development (LDRD) program.

\bibliography{neon}

\begin{thebibliography}{10}
\bibitem{neon_scint_corr}
The results from~\cite{neon_scint} have been renormalized to a new measurement
  of scintillation yield in helium~\cite{lenou_taup}.
\end{thebibliography}

\end{document}